\documentclass[11pt]{article}

\usepackage{graphics}
\usepackage{amsmath,amsfonts}

\usepackage{amsmath}
\usepackage{amssymb}
\usepackage{amsthm}
\usepackage{psfrag}
\usepackage{graphicx}
\usepackage{color}
\usepackage{mathtools}
\usepackage{hyperref}
\usepackage{array}
\usepackage{cite}

\newcommand{\beq}{\begin{equation}}
\newcommand{\eeq}{\end{equation}}
\newcommand{\beqa}{\begin{eqnarray}}
\newcommand{\eeqa}{\end{eqnarray}}
\newcommand{\bea}{\begin{eqnarray}}
\newcommand{\eea}{\end{eqnarray}}
\newcommand{\LL}{{\cal L}}
\newcommand{\OO}{{\cal O}}

\begin{document}

\begin{center}
\vspace{2.5cm}

\sc{\huge \bf 
Recipes for Oscillon Longevity
}

\vspace{1cm}

Jan Oll\' e$^{1}$, Oriol Pujol\`as$^{1}$ and Fabrizio Rompineve$^{2}$\\

\vspace{1cm}

{\it $^1$Institut de F\'{\i}sica d'Altes Energies (IFAE)\\ 
\normalsize\it The Barcelona Institute of  Science and Technology (BIST)\\
\normalsize\it Campus UAB, 08193 Bellaterra (Barcelona) Spain}\\
\bigskip
{\it $^2$ Institute of Cosmology, Department of Physics and Astronomy,\\
\normalsize\it Tufts University, Medford, MA 02155, USA}\\

\end{center}


\begin{abstract}
\noindent Oscillons are localized states of scalar fields sustained by self interactions. They decay by emitting classical radiation, but their lifetimes are surprisingly large. We revisit the reasons behind their longevity, aiming at how the  shape of the scalar potential $V(\phi)$ determines the lifetime.  The corpuscular picture, where the oscillon is identified with a bound state of a large number of field quanta, allows to understand lifetimes of order of $10^3$ cycles in generic potentials. At the non-perturbative level, two properties of the scalar potential can substantially boost the lifetime: the flattening of $V(\phi)$ and the positivity of $V''(\phi)$. These properties are realized in the axion monodromy family of potentials. Moreover, this class of models connects continuously with an exceptional potential that admits eternal oscillon solutions.
We check these results with a new fast-forward numerical method that allows to evolve in time to stages that cannot be otherwise simulated on a computer. The method exploits the attractor properties of the oscillons and fully accounts for nonlinearities. We find lifetimes up to $10^{14}$ cycles, but larger values are possible.
Our work shows that oscillons formed in the early Universe can be stable on cosmological time scales and thus contribute to the abundance of (ultra)light scalar dark matter.
\end{abstract}

\newpage 

\tableofcontents

\newpage

\section{Introduction}

Real scalar fields can exhibit a variety of dynamical phenomena in the early Universe, which makes them interesting and useful for inflationary and dark matter model building. Among these, a quite peculiar one can occur in the presence of attractive self interactions: the formation of localized, oscillating field configurations, known as \emph{oscillons}~\cite{Bogolyubsky:1976yu, Gleiser:1993pt, Copeland:1995fq} and also referred to as~\emph{axitons} when the scalar field potential is periodic, e.g.~in the QCD axion case~\cite{Kolb:1993hw, Kolb:1994fi}.

Since particle number is not a conserved quantity in a real scalar field theory, oscillons are necessarily subject to decay. In very weakly coupled theories, such as is the case of axions, this occurs very slowly via the classical radiation of scalar waves. Oscillon lifetimes then vary dramatically with the shape of the potential, and so does their observational impact. Understanding the causes of an oscillon's longevity is thus an important task, to which a lot of work has been devoted (see e.g.~\cite{Kasuya:2002zs, Fodor:2006zs, Saffin:2006yk, Fodor:2008es, Gleiser:2008ty, Fodor:2009kf, Amin:2010jq,  Hertzberg:2010yz, Salmi:2012ta, Andersen:2012wg, Saffin:2014yka, Mukaida:2016hwd, Ibe:2019vyo, Gleiser:2019rvw, Olle:2019kbo, Kawasaki:2019czd, Zhang:2020bec, Kawasaki:2020jnw, Zhang:2020ntm}).

In this paper, we approach this problem in two directions: First, we provide an analytical understanding behind the longevity of large-amplitude oscillons aiming to identify the properties of the scalar potential $V(\phi)$  that can enhance longevity. Secondly, we present a new numerical method by which one can reliably estimate the lifetime of an oscillon even when its evolution is too long to be entirely simulated on a computer. We thus obtain evidence of oscillons which can survive until today and constitute an interesting dark matter component.

Our analytical approach is anchored on the properties of an {\em exceptional} scalar potential,  $V \sim \phi^2 + \epsilon\phi^2\log\phi$,  that is almost quadratic across an exponentially large field range when $\epsilon\ll 1$. It is known that this potential leads to (classically) infinitely long lived oscillons, both in $1+1$ \cite{Dvali:2002fi} and $3+1$ dimensions \cite{Kawasaki:2015vga}.
This largely explains the longevity of oscillons supported by ``monodromy'' potentials~\cite{Silverstein:2008sg, McAllister:2008hb} that behave like some power-law $\phi^{2p}$ at large field values, and which we have shown in our previous work~\cite{Olle:2019kbo} to have the largest lifetimes ever computed before. 
Indeed, in the $p\to1$ limit, these potentials closely resemble the exceptional potential at large field values and it is then possible to analytically estimate the lifetimes of the associated oscillons by perturbing the exact solution. This shows that, in contrast to our previous expectations, monodromy oscillons with $1/2<p\lesssim 1$ can be extremely long-lived, much beyond what can be simulated on a computer.

Our analytical findings then motivate the search for a novel numerical strategy to probe such large lifetimes. Recently, a new method in this direction has been proposed~\cite{Ibe:2019vyo} (and applied to monodromy potentials in \cite{Kawasaki:2019czd}), which takes advantages of two key concepts. First, the time evolution can be {\em fast-forwarded} by means of a clever trick: rather than numerically evolving the whole field configuration in a discretized lattice, one just computes the emitted power in scalar radiation from a given oscillon-like configuration. A simple idea then is to just {\em assume} a certain profile with a single oscillation frequency, and obtain the energy loss rate $\Gamma(\omega)$ for an "oscillon" of that frequency. The result can then be easily integrated and gives an estimate of the lifetime. The second ingredient in~\cite{Ibe:2019vyo, Kawasaki:2019czd} was the computation of $\Gamma(\omega)$ by linearizing the field equation for the radiation field. This simplifies the computation at the expense of neglecting some nonlinearities which might be important. 

In this work, we propose what we believe is an improvement on this method by adopting the fast forward strategy of computing the energy loss $\Gamma(\omega)$ but keeping the full nonlinear equation (see also~\cite{Zhang:2020bec} for progress in this direction). Starting from quite random initial conditions, one can populate real oscillons in a range of frequencies by giving a sufficient (but realizable) relaxation time. Even in the most extreme cases when the lifetimes are very long, the relaxation time needed to see the trajectory of the true oscillon in the $\Gamma-\omega$ plane is much smaller than the actual lifetime. This strategy still  `buys' a lot of time, and for this reason we call the method {\em relax and fast forward}.

Our method, while more costly to implement, offers the great advantage of dealing always with the full nonlinear problem, basically with no assumptions on the oscillon shape and power spectrum at a given time. This exploits the attractor property of the true oscillon configuration, which drives many different initial conditions into the same oscillon (the same trajectory in the $\Gamma-\omega$ plane), in a rather short relaxation time.  

Equipped with our numerical strategy, we are able to extract oscillon lifetimes for any value of $p$ and confirm our analytical estimates. In particular, this allows us to complete our investigation of potentials with $p<0$. In contrast to our previous expectations~\cite{Olle:2019kbo}, we find that flattening the potential with $p<-1/2$ does not lead to increased lifetimes with respect to the case $p\lesssim 1$.

The rest of this paper is structured as follows: In Sec.~\ref{sec:understanding} we present qualitative criteria to obtain long-lived oscillons, while we provide analytical estimate for generic, exceptional and monodromy-like potentials in Secs.~\ref{sec:non-exceptional}, \ref{sec:exceptional} and \ref{sec:monodromy} respectively. We then introduce our numerical strategy in Sec.~\ref{sec:method} and present numerical results for oscillon lifetimes in Sec.~\ref{sec:results}. We offer a final discussion and conclusions in Sec.~\ref{sec:conclusions} and provide more details about our numerical method in the Appendix.

\section{Understanding Longevity}
\label{sec:understanding}

To set the stage, we consider  a single real scalar of mass $m$ 
with lagrangian $\LL = (\partial\phi)^2/2-V(\phi)$ and with a generic potential,
\begin{equation}\label{Vgeneric}
V \left(\phi\right) = \frac{1}{2} \, m^2 \phi^2  + \sum_{n=2}^{\infty}\, 
g_{2n} \; \phi^{2n}~,
\end{equation}
with a negative quartic coupling $g_4<0$.
For axions, it is convenient to write $V=m^2 \; F^2 \; v(\phi/F)$ with $v$ a dimensionless function $v(x)=x^2/2+\dots$ and with $F$ the axion decay constant. The quartic coupling is then naturally of order $g_4\sim (m/F)^2$ (in general, $g_{2n}\sim m^2/F^{(2n-2)}$) and the weak coupling regime maps to $m\ll F$. 

Despite the smallness of the couplings in the Lagrangian, a strong collective interaction arises when a large occupation number of axion scalar  quanta $N$ is considered, as can be understood by defining a \emph{collective coupling} $\lambda = N g_4$~\cite{Dvali:2017ruz}. For $N\sim 1/g_4\gg 1$, the system can be treated classically and in particular when $\lambda\sim 1$ one expects the formation of bound states held together by attractive self interactions, i.e.~oscillons. However, since particle number is not conserved for a real scalar field, such bound states are metastable. For $g_4<0$, higher dimensional operators in~\eqref{Vgeneric} need to be important at large field values, otherwise the potential would be unstable. These operators can then play an important role in determining the longevity of the scalar bound state. 

While the lifetime of oscillons depends crucially on the higher dimensional operators in $V(\phi)$, it is possible to qualitatively understand the features that can lead to large lifetimes in a model-independent (and non-perturbative) way. First, let us consider the quantity
\beq\label{Deltaw}
\Delta \omega \equiv m- \omega_{\text{osc}},
\eeq
where $\omega_{\text{osc}}$ is the frequency of oscillations of the field $\phi$ in the oscillon configuration. The quantity $\Delta\omega$ can be thought of as the binding energy per particle: each corpuscle decreases the total energy of a set of $N$ quanta by $\Delta\omega$ by being in the lump, therefore the very existence of an oscillon requires a significant $\Delta \omega$. While the dependence of $\Delta\omega$ on $V(\phi)$ may be complicated, in the classical field theory limit with $N\sim 1/g_{4}\gg 1$, it is very reasonable to expect that this quantity is directly controlled by the deviation of $V(\phi)$ from the free part in the oscillon core
\begin{equation}
\Delta V\left(\phi\right) \equiv \frac12  m^2 \phi^2 - V(\phi).
\label{DeltaV}
\end{equation}
We will refer to $\Delta V$ as the \emph{binding potential}: in order to form and persist, an oscillon should be characterized by a significant $\Delta V\left(\phi\right)$ in its core. This in turn leads to a rough criterion to estimate the oscillation amplitude in the core, $\phi_0$, for any given $V(\phi)$, i.e. 
\beq\label{crit}
\Delta V(\phi_0) = \delta \; \frac{m^2\,\phi_0^2}{2}
\eeq
with $\delta$ a sizeable fraction of 1.\footnote{This criterion is actually supported by our numerical studies with monodromy-like potentials, as we shall see below. For instance, we show in Fig.~\ref{fig:delta-w} the correlation between $\Delta \omega$ and $\delta$.
Typical values in simulations for different potentials give a range of $\delta$ from $0.2$ to $0.85$. This supports the criterion suggested above that $\delta$ must be a sizeable fraction of unity. Moreover, another result from Fig.~\ref{fig:delta-w} is that in the $p\to1$ limit, $\delta$ is of order $0.15-0.2$.}

Secondly, once an oscillon has formed, its lifetime is dictated by its (in)efficiency to radiate classical scalar waves. In this respect, it is well known that fluctuations of an homogeneous oscillating field can undergo resonant enhancement. The same can happen in the localized oscillon configuration, i.e. the amplitude of modes whose wavelength is smaller than the oscillon size can be potentially enhanced. When this happens, the oscillon configuration is quickly disrupted. In analogy with the homogeneous case, the efficiency of such a resonant enhancement is controlled by $V''(\phi_{\text{osc}}(t,r))$, which is on general grounds controlled by the effective mass $V''(\phi)$. In particular, we expect that potentials with negative $V''(\phi)$ for some large enough field values will be characterized by efficient resonant decay of oscillon configurations. The same logic  suggests that, generically, negative but small effective mass,  $|V''(\phi)|/m^2 \ll 1$, at the core can be `tolerated' because both the length- and time- scales of the oscillons are of order $\sim 1/m$, so the naively tachyonic instability can be inefficient.

In sum, a qualitative criterion for the presence of long-lived oscillons is that the potential $V(\phi)$ satisfies these 2 conditions: 
\begin{itemize}
\item[$\bf{1)}$]{$\Delta V(\phi)$ is maximized.}
\item[$\bf{2)}$]{$V''(\phi)$ (including the sign) is maximized.}
\end{itemize}

Interestingly, the two conditions above are somewhat antagonistic. For instance the obvious way to increase the binding potential $\Delta V(\phi)$ is by ``bending'' $V(\phi)$ so that it decreases with $\phi$ past some value. However, this means that there is a maximum, with $V''(\phi)<0$, so that radiation would be efficient in this case. In turn, requiring that $V''(\phi)$ is bounded from below limits the size of the binding potential $\Delta V(\phi)$.

At first, it is not immediate to tell which of the conditions above is more important, nor what combination of the two should one optimize  in order to maximize the lifetime. It is rather obvious that plateau-shaped potentials are good at satisfying $\bf{1)}$, while potentials very close to  $\phi^2$  are good at satisfying $\bf{2)}$. 

It turns out that condition $\bf{2)}$ plays a stronger role in enhancing longevity. 
One way to understand this is that even if $V(\phi)$ deviates from $m^2\phi^2/2$ only slowly, one can still find a large enough amplitude $\phi_0$ such that condition \eqref{crit} (sizeable binding potential) is met. If this happens while the potential is still close to quadratic (therefore $V''(\phi)/m^2$ close to $1$) then one does not expect much radiation, which therefore leads to a long lifetime. 

This logic is further confirmed by the fact that the potential $V=-\phi^2 \log\phi$ (which incarnates precisely the limit of being close to a quadratic potential) turns out to admit {\em  exact} non-dissipative oscillons, as we discuss below (see Sec.~\ref{sec:exceptional}).
Oscillons in close-to-quadratic potentials, then, are expected to have a boosted lifetime. 

The oscillons in $V=-\phi^2 \log\phi$ are classically eternal for any amplitude. However, this potential displays a maximum at high enough $\phi$ (with order-1 negative $V''$), which seems to go against condition $\bf{2)}$ above. The crack in the argument is that the oscillons that explore the maximum in fact decay because they have unstable resonant modes \cite{Ibe:2019lzv}. Therefore our logic still applies. A similar thing happens with breathers in 1+1 sine-Gordon theory: at the quantum level the spectrum of amplitudes is quantized \cite{Vachaspati:2006zz}, and the `ground state' breather does not probe the maximum. 

We now discuss separately the longevity for the 3 relevant types of potentials: of generic form, exceptional form, and finally we go to the family of monodromy potentials.

\subsection{Lifetime estimates}
\label{sec:non-exceptional}

In this section we follow the corpuscular description of bounds states \cite{Dvali:2012en,Dvali:2013eja} (devised to understand generic localized objects like black holes or solitons) to obtain estimates of the oscillon lifetime in terms of the couplings appearing in the potential. The starting point of the corpuscular picture is that classical field solutions can be re-interpreted as mean-field descriptions of the (bosonic) quantum field in the Bose-Einstein condensate limit, that is when there is a large occupation number of the same state. This picture is well suited for oscillons when the couplings in the Lagrangian are small. First, this is because in a  weakly coupled theory it makes sense to view the bound states as being composed of approximately free field quanta. 
Second, the oscillon mass typically scales like $E\sim m / g_4 \gg m$ so indeed one can view it as composed of a large number $N\sim 1/g_4$ of quanta, which is consistent with the large occupation number picture.

Since the $N$ quanta basically occupy the same state, one can identify the wavelength of the  occupied quantum state with the oscillon radius, $R$, which is around $1/m$, but often slightly bigger. A heuristic way to obtain $R$ is to picture that the scalar field quanta are trapped (dynamically) inside a spherical box of radius $R$. The lowest energy modes should then exhibit a  relation between the oscillation frequency and the radius of the form
\beq\label{wo}
\omega=\sqrt{m^2-R^{-2}}~.
\eeq
Since the oscillation frequency indeed satisfies $\omega<m$, one can  use this equation as a definition for the oscillon radius (the numerical simulations show that this is a good estimate for the radius) once the oscillon frequency is `measured'.

The departure of $\omega$ from $m$ is an important property of oscillons, as it leads to a notion of binding energy $\Delta \omega$ according to
\beq\label{omega}
\omega_{osc}=m-\Delta \omega~.
\eeq
This $\Delta \omega$ represents the gain in energy (per quantum) in forming the localized bound state. In numerical simulations,  $\Delta \omega/m$  is measured to be in the $10^{-2} - 10^{-1}$ range.\footnote{The lower end of the window can be understood from the requirement of stability of the oscillon in the non-relativistic approximation, see \cite{Mukaida:2016hwd,Kawasaki:2019czd}.}
For small binding energy $\Delta\omega/m \ll1$, one obtains the usual non-relativistic relation between $\Delta \omega$ and the momentum $1/R$,
\beq\label{NRwR}
\Delta \omega \simeq \frac{1}{2 m R^2} ~.
\eeq
With this corpuscular picture in mind we are now ready to estimate the oscillon lifetime from the form of the potential~\eqref{Vgeneric}. 
We are now viewing the oscillon as a finite density concentration of $N\sim 1/g_4$ particles, therefore one can attempt to estimate the lifetime by means of the usual formula
\beq\label{rateEstimate}
\Gamma \sim \,v \,\sigma \,n,
\eeq
where $\sigma$ is the scattering cross section associated to the processes that generate radiation, 
$n$ is the number density and  $v$ the typical velocity of particles. The number density is
$$
n= \frac{N}{V}  \sim \frac{1}{g_4 R^3}
$$ 
with $R$ the typical oscillon size, which is related to the frequency of oscillation.  
The typical speed of the quanta in the soliton can be estimated as $v\sim p/E$ with  $E=\omega$ and $p\sim 1/R$. Using \eqref{omega}, this 
reduces to approximately $1/(mR)$ for small $\Delta\omega$.
The cross section for individual $3\to 1$ conversion processes from the quartic coupling is estimated as $\sim g_4^2 m^{-2}$.
Since there are $~1/g_4$ quanta in the bound state, the total cross section is enhanced by a $1/g_4$ factor. Collecting all terms, one arrives at
\beq\label{GammaEst}
\Gamma \sim \frac{1}{g_4 R^3} \,\frac{1}{Rm} \,\frac{g_4}{ m^{2}} = \left( \frac{\Delta\omega}{m} \right)^2 \; m
\eeq
where we used \eqref{NRwR} in the last step.

There are 3 important features of this estimate: 
First, oscillons in simplest potentials (including the sinusoidal, which is relevant for the QCD axion) exhibit a rather small binding energy (per quantum) 
$$\frac{\Delta\omega}{ m} \sim 10^{-2}$$
and this according to \eqref{GammaEst} leads to a considerably long lifetime, of the order of $10^3$ oscillations. This naive estimate actually matches the lifetime which is determined by numerical computations for quartic and sinusoidal potentials. 

Second, \eqref{GammaEst} does not depend on the magnitude of the coupling $g_4$, as expected. In the $g_4\ll1$ limit, where the dynamics simplifies to mean field theory, the field can always be rescaled at will in order to fix the magnitude of {\em one} coupling, which we can take to be $g_4$. This also makes manifest that oscillon properties such as its binding energy $\Delta\omega$ or its lifetime must depend exclusively on the set of higher order self-interactions, $g_n$ with $n\geq 6$.

Third, the estimate \eqref{GammaEst} is based on looking only at the quartic self-coupling so this  implicitly assumes that higher order couplings, $g_n$ with $n\geq6$, do not give rise to cancellations that would significantly change the estimate. This is what we mean by  {\em non-exceptional} potentials. Note that {\em all} couplings $n\geq6$ contribute to processes where the emitted particle has energy $\sim 3m$, which can compete with the $3\to1$ channel from $g_4$, therefore it is conceivable that higher order couplings may affect the rate. In the introduction of Sec.~\ref{sec:understanding}, we discussed in an intuitive way the possible ways by which the final rate can be suppressed, based on the shape of $V(\phi)$ (and $V''(\phi)$). In the next subsections we show this more explicitly.

By the same logic, one expects that lifetimes (identified with $1/\Gamma$) substantially differing from \eqref{GammaEst} should correspond to the situation where there is destructive interference between different channels (from different couplings). This should also translate into having enhanced emitted power in the different multiples of the fundamental frequency ($3 w_{\text{osc}}$, $5 w_{\text{osc}}$, etc). We shall not show results for the power spectra of the radiation from oscillons, but we have checked that this is indeed the case: for potentials with longer lived oscillons, the power spectra in higher harmonics become more comparable, which further confirms this picture.

Finally, let us take a  brief detour as we want to emphasize that the kind of reasoning presented here is strictly parallel to the way of understanding a more familiar yet nontrivial bound state appearing in a very simple theory: namely, positronium in QED. Even though in this case we are dealing with fermions, and there are only two particles in the bound state, the same use of \eqref{rateEstimate} also allows to compute the positronium decay rate. As is well known, the decay rate differs for the two spin states. The singlet state (para-positronium) decays by emitting 2 photons ($\sigma \sim \alpha^2$), while the triplet (ortho-positronium) emits 3 photons ($\sigma \sim \alpha^3$). The rate picks 3 more powers of $\alpha$ from the number density which is  obtained from the Bohr radius $1/(m_e\alpha)$, reproducing the usual values of the lifetime $\alpha^5 m_e$ or $\alpha^6 m_e$ for (para- or ortho-) positronium. Perhaps this makes one more confident with the estimate \eqref{GammaEst} above.

\subsection{Exceptional potential}
\label{sec:exceptional}

Let us now discuss an exceptional potential that escapes the logic presented in the previous section. It consists in the special form ``$V_{\log}(\phi)"$$\propto \phi^2 + \phi^2 \log\phi$. Heuristically, it is somewhat unsurprising that this potential leads to non-radiative localized solutions because the effective mass-squared at the origin blows up and so it is energetically impossible to emit scalar radiation. Still, this potential stores an additional surprise: the resulting oscillon solutions can be obtained analytically. 

For later use, we perform an arbitrary rescaling and introduce a parameter $\epsilon$ to write the potential in the form
\begin{equation}\label{Vlog}
V_{\log}\left(\phi\right) =  \frac{1}{2} \, (1+\epsilon) m^2 \phi^2  - \frac{1}{2} \epsilon\; m^2 \phi^2\, \log\left(\frac{\phi^2}{F^2}\right) ~.
\end{equation}
This potential has 2 somewhat `unwanted' features: i) it is not exactly analytic around $\phi$ due to the $\log\phi$; ii) it is unbounded from below -- clearly at $\phi \geq e^{\frac{1}{2}(1+\frac{1}{\epsilon})} F$ it becomes negative.

The potential \eqref{Vlog} leads to an equation of motion of the form
\beq
\frac{\partial_r^2 \phi}{\phi} + \frac2r\,\frac{\partial_r\phi}{\phi}  - \frac{\partial_t^2\phi}{\phi}= \frac{V_{\log}
'(\phi)}{ \phi} = m^2 - \epsilon m^2 \log(\phi^2/F^2)
\eeq
Despite being nonlinear, this equation is factorizable \cite{Dvali:2002fi,Kawasaki:2015vga} -- which is at the very root of the exceptional properties of $V_{\log}$. Indeed, the factorized ansatz
\beq\label{exact}
\phi_{i}(t,r) = A(t) B(r) 
\eeq
leads to an immortal localized solutions with a Gaussian profile
\beq\label{gauss}
B(r)= e^{-(r/R)^2} \qquad { \rm with } 
\, \qquad R=\sqrt{\frac2\epsilon}\,\frac1m
\eeq
provided that the overall amplitude $A$ satisfies the ODE 
$$
\ddot A =  -m^2\,A\;\left(\,1 + 3\epsilon \,  - \,  \epsilon \;\log(A^2/F^2)\;\right)~,
$$
that is, it oscillates according to same logarithmic potential. Since the equation above can be solved by quadratures, the evolution of $A(t)$ is periodic, therefore this is a non-radiating ever-lasting oscillon. 

Two comments are in order. First, the potential $\phi^2\log\phi$ is encountered in some supersymmetric models (see e.g.~\cite{Dvali:2002fi} and~\cite{vonHarling:2019gme} in a different context). Second, the mere fact that the theory admits exact stable localized excitations suggests that this might be due to some kind of integrability property. Indeed the potential $\phi^2\log\phi$ is a member of a class of potentials with the property \cite{Minahan:2000ff} that they contain static localized `lump' field configurations, whose stability leads to a reflectionless Schr\"odinger potential \cite{Minahan:2000ff,Zwiebach:2000dk,Cooper:1994eh}. Interestingly, for the $\phi^2\log\phi$ potential the `lump' (the above solution with $A=e^{1/2\epsilon}F$) leads to a Schr\"odinger problem with a purely quadratic potential, so it has an infinite number of evenly-spaced bound states.

\subsection{{\em Monodromy} potentials}
\label{sec:monodromy}
We now consider the 1-parameter family of potentials:
\begin{equation}\label{eq:potential}
V\left(\phi\right) = \frac{1}{2 p} \, m^2 F^2 \,\left[ \left( 1 + \frac{\phi^2}{F^2} \right)^p - 1 \right] 
= \frac{m^2\phi^2}{2} - \frac{1-p}{4}\,\frac{m^2}{F^2} \phi^4  + \dots,
\end{equation}
which was our focus in~\cite{Olle:2019kbo}.
Oscillons form for $p<1$, which is when both $\bf{1.}$ the quartic coupling is negative and $\bf{2.}$ the potential is smaller than the free part, $V(\phi) < m^2 \phi^2 / 2 $. It is useful to analyze certain limiting cases of this class of potentials.
\begin{itemize}

\item[{\bf I)}] $\mathbf{p\to -\infty}$

The potential becomes
\begin{equation}\label{Vtilde}
V\left(\phi\right) 
\simeq \frac{m^2 \tilde F^2}{2} \left( 1- e^{-\phi^2 / {\tilde F}^2} \right) + \OO\left(\frac{1}{p}\right)
\end{equation}
with $\tilde F =  F / \sqrt{1-p}$. Therefore, at large negative $p$ the potential is insensitive to the value of $p$ as should be the oscillon properties. 

\item[{\bf II)}]  $\mathbf{p\to 1} $ 

Expanding around $p=1$, the potential becomes 

\begin{equation}\label{Vsmallp-1}
V\left(\phi\right) \simeq \frac{1}{2} \, m^2 \left\{ \phi^2  - (1-p) \Biggr[ \left(F^2+\phi^2 \right) \log\left( 1+\frac{\phi^2}{F^2} \right)  \; - \phi^2 \Biggr] + \dots \; \right\}
\end{equation}
the dots denoting corrections of order  $\OO\left[  (1-p)^2 \log^2( 1+\phi^2/F^2) \right] $. This potential is of course smooth at the origin, where $V=m^2 \phi^2 /2 - \dots$, whereas at large field values it is very similar to the exceptional potential \eqref{Vlog}. Indeed, one recovers \eqref{Vsmallp-1} by replacing $\phi^2\to\phi^2+F^2$ in the $\phi^2\log\phi$ term of \eqref{Vlog}, with the identification 
\beq\label{eps}
\epsilon\equiv 1-p \ll1~.
\eeq

Given that the series expansion near $p=1$ is really an expansion in $(1-p) \log\phi$, the potential \eqref{eq:potential} can be safely truncated as \eqref{Vsmallp-1} in a rather large range in field space, $\phi \lesssim \tilde{\phi}$, with 
\beq\label{phi*}
\tilde{\phi} \equiv e^{\frac{1}{ 2 \epsilon}}  \;F \gg F~.
\eeq
Therefore, in the $1-p\ll1$ limit \eqref{eq:potential} reduces to the exceptional potential $V_{\log}$ \eqref{Vlog} in a large range of field space, except for small values $\phi\lesssim F$.

As explained above, the field  amplitude $\phi_0$ in the oscillon core must be large enough to have a significant binding potential, expressed as the condition \eqref{crit} with $\delta$ of order of, say, $\sim 1/10$. Applied to the potential~\eqref{Vsmallp-1}, this condition gives 
\beq\label{phi0delta}
\phi_0\simeq e^{\frac{\delta}{2\epsilon} + \frac12} \;F
\eeq
which in the $\epsilon \ll1$ limit is large but smaller than $\tilde{\phi}\sim \phi_0^{1/\delta}$ (in units of $F$).

\begin{figure}[t]
\centering
\includegraphics[width=0.6\textwidth]{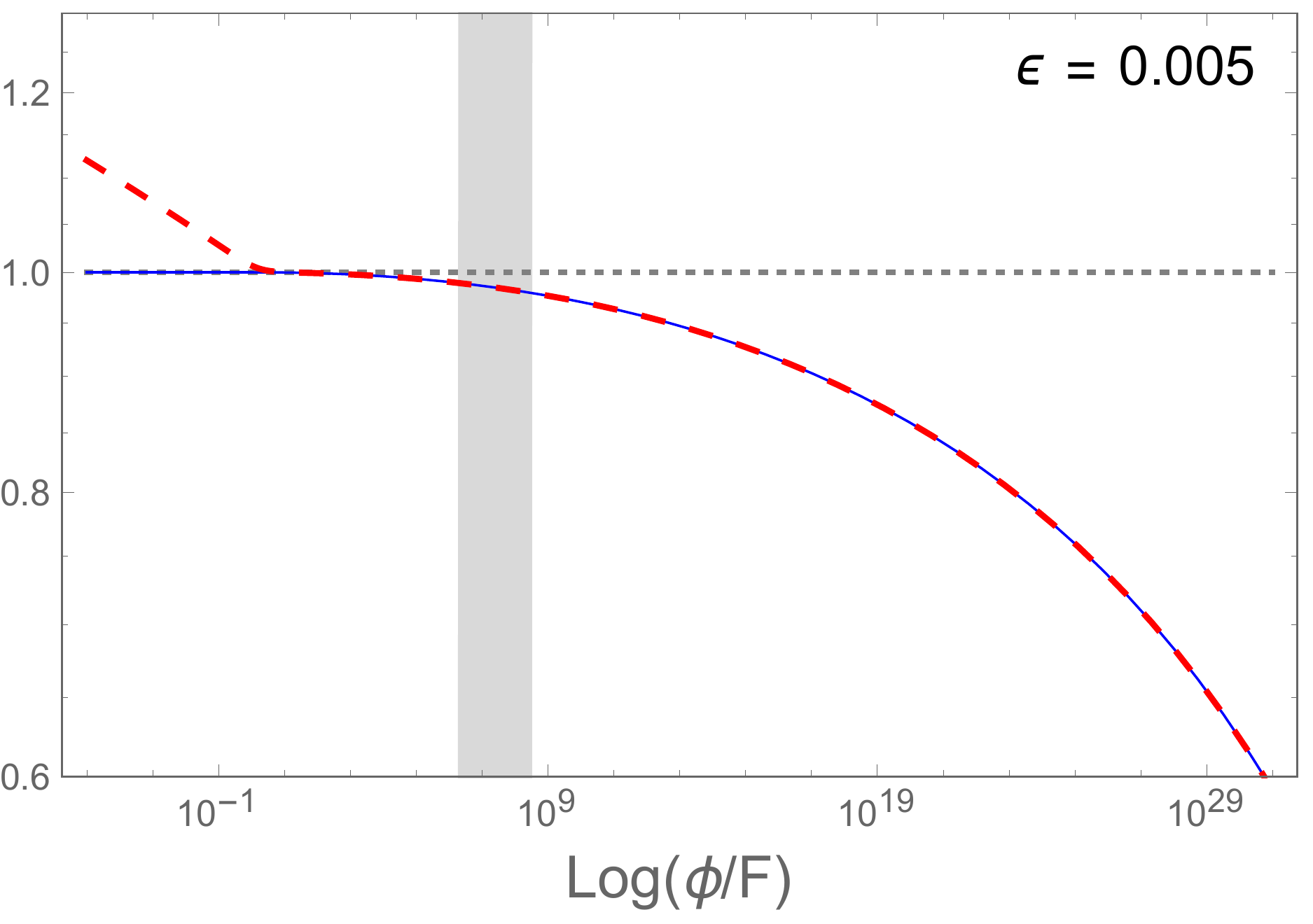}
\caption{\small
Ratio of the truncated potential \eqref{Vsmallp-1} to the full monodromy potential \eqref{eq:potential} (blue solid line) and of the exceptional  \eqref{Vlog} to the monodromy potentiald (red dashed line) as a function of $\phi$, for $\epsilon=0.005$. Clearly, there is a \eqref{Vsmallp-1},  \eqref{Vlog}  are very similar (and close to \eqref{eq:potential}) for many orders of magnitude. The vertical band represents the value of $\phi_0\sim e^{\delta/2\epsilon}$.}
\label{fig:potentials}
\end{figure}

In sum, in the $\epsilon \equiv 1-p \ll 1$ limit,  the approximation \eqref{Vsmallp-1} is well justified and the oscillon amplitude is expected to lie in the  window
$$
F\ll \phi \ll \tilde{\phi}~
$$
where the potential is very close to the exceptional one \eqref{Vlog}, see Fig.~\ref{fig:potentials}. This class of potentials is then expected to lead to oscillons with enhanced longevity is the limit $p\rightarrow 1$. 

\subsection*{Scaling of the lifetime for $p\to1$}

The expectation above is confirmed by numerical simulations, which we will present in Sec.~\ref{sec:numerics}. In particular, when $p$ is very close to $1$, oscillon lifetimes increase seemingly without bound, and certainly to values that cannot be simulated in present day computers. 

In the limit $\epsilon = 1-p \to1$, however, we can exploit the similarity of the monodromy potential with the form $\sim \phi^2 \log\phi$. 
Indeed,  $V\simeq V_{\log}$ for a very large range in field space  $F\ll \phi \ll \tilde{\phi}$ in the $\epsilon\ll1$ limit. The oscillon lifetime then can be estimated by assuming that oscillons supported by~\eqref{eq:potential} should be very close to the `immortal' oscillons \eqref{exact}.

This suggests to split the solution as
\beq\label{split}
\phi=\phi_{i}+\delta \phi
\eeq
where $\phi_i$ is the `immortal' solution \eqref{exact} and we assume $\delta\phi$ is a small deviation from it in a large space-time region.\footnote{Of course, this assumption must break down asymptotically far away  but still it is possible to see that  there is a `radiation zone' where $\delta\phi$ is small and takes energy from the oscillon core.} Plugging \eqref{split} into the equation of motion one arrives at
\beq\label{deltaphi}
\Box \delta\phi + V''(\phi_i(t,r))\; \delta\phi =-\left( \Box \phi_i +  V'(\phi_i) \right) \equiv j(t,r)
\eeq
where terms of order $\delta\phi^2$ have been  dropped. 
Using the equation of motion that $\phi_i$ satisfies, the source $j(t,r)$ is identified as
\beq\label{source}
 j(t,r) = -\left( V'-V'_{\log} \right)\Big|_{\phi=\phi_i}=
  \epsilon\,m^2 \;\phi  \left[ \log\left(1+\frac{\phi^2}{F^2}\right)  -  \log\left(\frac{\phi^2}{F^2}\right) \right] \Biggr|_{\phi=\phi_i}
\eeq
The function $V'_{\log}(\phi)-V'(\phi)$ has a maximum at around $\phi\simeq 0.5 \,F$, where it takes the value $\simeq 0.8\, \epsilon\, m^2\, F$. For small field values, $\phi\ll F$, it goes like $\sim -2m^2\epsilon \phi \log(\phi/F)$, and at large $\phi$ it decays like $\sim \epsilon~m^2\;F^2/\phi$.
Having in mind large amplitude oscillons with $\phi_{0} \sim  e^{\frac{\delta}{2\epsilon} + \frac12} \;F $ as given in \eqref{phi0delta}, this implies that the source term $j$ is tiny.  Compared to $V'$ in the full solution, it is suppressed by a factor (neglecting logarithms)
\beq\label{suppression}
\frac{|\,V'-V'_{\log} \,|}{V'} \lesssim \frac{F^2}{\phi_0^2} \sim  e^{- \frac{\delta}{\epsilon}} ~.
\eeq
While the characteristic size of the oscillon (where, say, most of the energy is concentrated) is  $R\simeq \sqrt{2/\epsilon}~m^{-1}$,  the Gaussian approximation holds up to much larger distances. 
With the amplitude in the core $\phi_0 \sim e^{\delta/2\epsilon}F$, one expects the Gaussian profile to hold  all the way up to $\phi\sim F$. Let us denote by $\tilde{R}$ the radius up to which the Gaussian profile is expected to hold, which is when $\phi\sim F$. 
Since at the origin the amplitude is \eqref{phi0delta} and we have the Gaussian profile, this is simply obtained from the condition $\exp\left(\frac{\delta}{2\epsilon} - \frac{\tilde{R}^2}{R^2}\right)\sim 1$, 
that is
\beq\label{R*}
\tilde{R} \sim \sqrt{\frac{\delta}{\epsilon}} R = \frac{\sqrt{\delta}}{\epsilon\,m}~.
\eeq
Note from the definition of $j(t,r)$ that this radius is where $j$ is expected to be localized (a shell around the oscillon considerably far from its core radius $R$). 

This relation also clarifies the main condition for the validity of the approximation \eqref{split}: a well defined approximately Gaussian-shaped oscillon profile requires that there is a clear separation between $R$ and $\tilde{R}$. This translates into the condition
\beq\label{epsilonlimit}
\epsilon \ll \delta~.
\eeq
More quantitatively, a factor $10$ in $\tilde{R}/R$ translates into 
$$
\epsilon \sim 10^{-2} \delta
$$
Numerically one finds that $\delta$ is around $0.2$ for $\epsilon\to0$ (see Sec.~\ref{sec:results}), therefore the Gaussian approximation should be reliable for $\epsilon \lesssim {\rm few}\; 10^{-3}$. 

Let us then proceed to estimate the lifetime of an oscillon in this regime. The effective scalar mass at the core $V''(\phi(t,0))$ is around $\simeq m^2 (1-\delta - 3 \epsilon)$ (with spikes going up to $m^2$ when $|\phi | \lesssim F $). In the crudest approximation one can neglect the space- and time- variation in \eqref{deltaphi}, and estimate the radiation from the massive KG equation with a source  \eqref{source}. Barring the potential resonant effects from the time dependent mass and source, one can then estimate the size of $\delta \phi$ upon  integration of \eqref{deltaphi} as $\delta\phi\simeq j /m^2$ . The emitted power can then estimated as $\dot{E}=4\pi r^2 T^{0r}\simeq \tilde{R}^2 \partial_r \delta\phi\,\partial_t \delta\phi\sim \tilde{R}^2 \partial_r j \, \partial_t j / m^4$. Recalling that the maximal magnitude is $j \lesssim \epsilon F^2 m^2/\phi \sim  \epsilon F m^2$ and using $\partial_t\sim \omega_{osc}$, $\partial_r\sim 1/\tilde{R}$, we arrive at
the estimate
\beq\label{Edot}
\dot E\sim \tilde{R}^2 \frac{w_{osc}}{\tilde{R}} \epsilon^2 F^2 \sim \sqrt\delta \epsilon F^2~.
\eeq
On the other hand, the oscillon energy can be estimated as
\beq\label{E}
E\sim R^3 m^2 \phi_0^2 \sim \epsilon^{-3/2} e^{\frac{\delta}{\epsilon}} \frac{F^2}{m}~.
\eeq

Combining \eqref{Edot} and \eqref{E} we arrive at the following estimate for the decay rate,
\beq\label{scaling}
\Gamma= \frac{\dot E}{E} \;\sim \;\epsilon^2\;\frac{F^2}{\phi_0^2} \;\frac{w_{osc}\,\tilde{R}}{m^2\,R^3}  \;\sim \;\epsilon^{\frac52} \;e^{-\frac{\delta}{\epsilon}} \; \sqrt{\delta} \; m
\eeq
where we used $ R  \sim m^{-1} / \sqrt\epsilon$, and $\omega_{osc}\sim m$. We thus find an exponentially large lifetime as $\epsilon\rightarrow 0$, i.e. $p\rightarrow 1$. In Sec.~\ref{sec:results} we will see that this simple formula correctly captures the magnitude and  behavior of oscillon lifetimes. 

Until now we have neglected gravity in our calculations. This is only valid if the oscillon is not particularly compact, i.e. as long as
\begin{equation}
\frac{R_S}{R} = \frac{2 G M}{R}\sim \epsilon^{-1} e^{\frac{\delta}{\epsilon}} \frac{F^2}{M_P^2} \ll 1.
\end{equation}
This translates into the upper bound $F\ll \sqrt{\epsilon}~e^{-\delta/(2\epsilon)}M_p$, which is particularly strong for $\epsilon\ll 0.01$. However, for phenomenological applications an even stronger bound on $F$ arises from dark matter overproduction, since the typical initial amplitude of the scalar field background required to form such oscillons also grows exponentially with $\epsilon^{-1}$. This latter constraint is discussed in Sec.~\ref{sec:conclusions}.

\end{itemize}

\subsection{Floquet analysis}

\begin{figure}[t]
\centering
\includegraphics[width=0.5\textwidth]{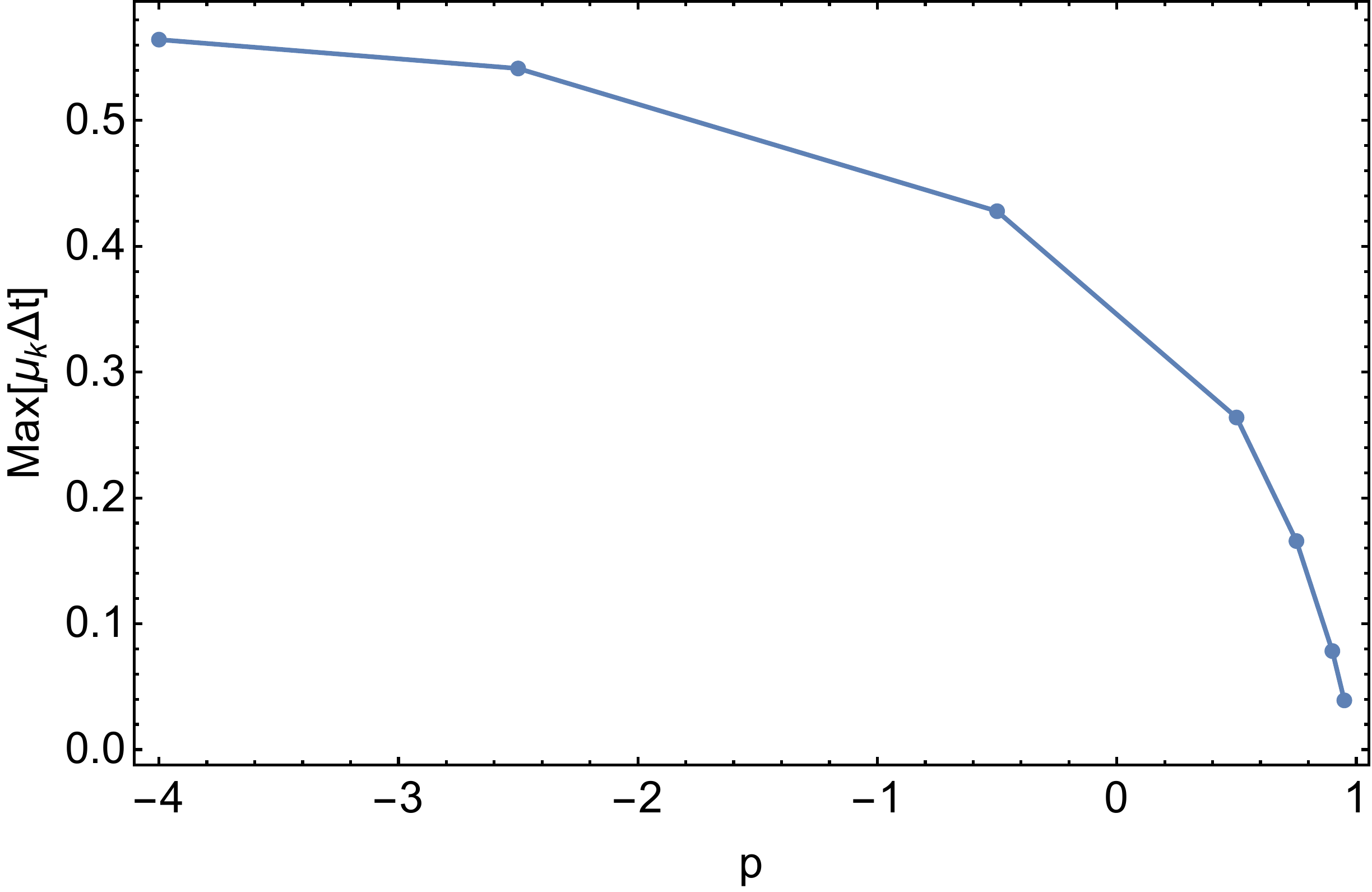}
\caption{\small Maximal values of the resonance efficiency parameter $\mu_k \Delta t$ as functions of $p$ for monodromy potentials.}
\label{fig:floquet}
\end{figure}

As argued at the beginning of this section, one of the conditions to ensure that oscillons can have long lifetimes is that resonant enhancement of short wavelength modes is suppressed. The aim of this subsection is to argue qualitatively that this is indeed the case for the monodromy potentials introduced in Sec~\ref{sec:monodromy}. 

We start by considering the Floquet diagram of an homogeneous oscillating scalar field with potential \eqref{eq:potential}. Some examples of such diagrams have been presented in our previous work~\cite{Olle:2019kbo}. Their crucial feature is the existence of a broad resonance band at $k\lesssim m$, whereas only very narrow bands arise for $k>m$. The distance in $k$-space between these bands increases with $p$: in particular bands are denser for $p<0$ than they are for $0<p<1$. Similarly, the values of the Floquet exponent $\mu_{k}$ are largest in the broad resonance band and increase with decreasing $p$, until they reach an approximately constant value for $p\ll -1$.

In analogy with the homogeneous case, we can understand the relevance of resonant enhancement for oscillons by considering the approximately spherical region of space $r\lesssim R$. Inside this region the localized field configuration is oscillating with amplitude $\phi_0>\phi_{\text{out}}$, where $\phi_{\text{out}}$ is the amplitude of the field outside the lump (here we neglect the dilution of this amplitude due to Hubble friction). The precise relation between $\phi_0$ and $\phi_{\text{out}}$ can be obtained numerically; for monodromy potentials \eqref{eq:potential} one finds that $\phi_0\simeq 3\phi_{\text{out}}$ during most of the oscillon lifetimes. Of course, only modes with $k\gtrsim R^{-1}$ can be enhanced, since those with $k< R^{-1}$ have wavelengths larger than the oscillon size, thus they cannot notice the presence of localized oscillations. Therefore, this implies that only modes with $R^{-1} \lesssim k\lesssim m$ can potentially undergo significant resonance. 

Additionally, those modes are actually enhanced only while they remain inside the oscillon configuration. Each mode propagates through space with a phase velocity $v=\nu/k=\sqrt{m^2+k^2}/k$, thus it exits the lump after a time $\Delta t\sim R/v\sim R~k/\sqrt{m^2+k^2}$. Therefore, modes are actually strongly enhanced only if $r=\mu_{k}~\Delta t\gtrsim 1$.\footnote{Very much like in the early Universe parametric resonance is effective only if $\mu_{k}H^{-1}\gtrsim 1$.} Since the oscillon size $R$ depends on $\phi_0$ (through $\epsilon$ in the Gaussian approximation considered above), one can compute values of $\mu_{k}~\Delta t$ as a function of $\phi_0$ and $k$ by means of standard Floquet methods, and extract its maximal size to understand the behavior of resonance with $p$.

The result of this analysis is shown in Fig.~\ref{fig:floquet} for some representative values of $p$, for which we have also numerically computed $R$ as a function of $\phi_0$. One can appreciate that $\mu_{k}~\Delta t$ is always smaller than $1$ for any value of $-4<p<1$. This suggests that resonance does not threaten the longevity of oscillons supported by the potentials \eqref{eq:potential} for those values of $p$, as confirmed by the numerical simulations which we present in the next section. Interestingly, resonance is shut off for $p\to 1$ and becomes more relevant for $p\ll -1$: as we will see, oscillon lifetimes will follow the opposite trend, with a notable exception around $p\simeq -1/2$.

\section{Numerical Simulations}
\label{sec:numerics}

We now turn to a numerical investigation of oscillon lifetimes. We start by presenting a novel numerical strategy, then present results for different values of $p$.

\subsection{Relax and Fast-Forward}
\label{sec:method}

\begin{figure}[t]
\centering
\includegraphics[width=0.7\textwidth]{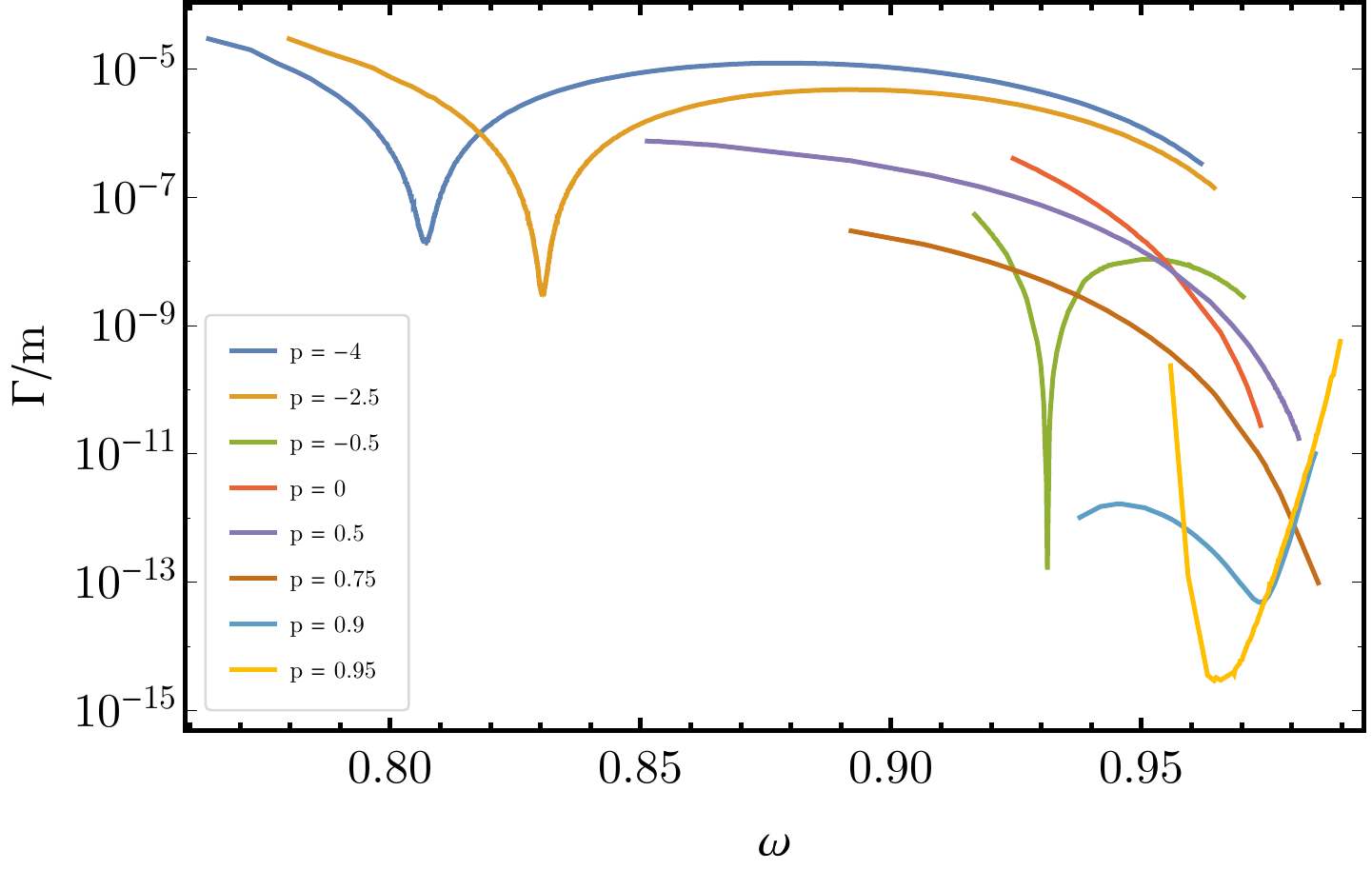}
\caption{\small
Numerical results for the decay rate $\Gamma(\omega)$, obtained using the relax and fast-forward method, for various values of $p$.}
\label{fig:Gamma-w}
\end{figure}

In this section we focus on a numerical method to compute $\Gamma = \dot{E}/E$, which then controls the lifetime $\tau$ of oscillons. It is convenient to treat $\Gamma$ as a function of $\omega$ so that we can write the differential equation
\begin{equation}
    \Gamma(\omega) = \frac{\dot{E}}{E} = \frac{1}{E} \left( \frac{dE}{d \omega}\right) \frac{d \omega}{dt},
\end{equation}
which can be easily inverted to give
\begin{equation}
    \tau = \int_{t_0}^{t_f} dt  = \int_{\omega_0}^{\omega_f} \frac{d\omega}{\Gamma(\omega)} ~ \frac{1}{E(\omega)} \left( \frac{dE}{d \omega}\right), \label{eq:tau_general_eqn}
\end{equation}
where $\omega_{0,f} \equiv \omega(t_{0,f})$ are some particular initial and final values for the frequency. In general, this equation can be numerically integrated to obtain the lifetime, and this is what we do in practice. However, as we will see in Fig.~\ref{fig:Gamma-w}, some values of $p$ show a pronounced dip in the function $\Gamma(\omega)$, in which both $E$ and $dE/d\omega$ are constants to very good approximation. Calling $\Gamma_*$ the minimal value of $\Gamma(\omega)$ in such dips, we can then write
\begin{equation}
    \Gamma (\omega) \approx \Gamma_* \left[ 1 + \frac{\beta}{\omega_*^2} ~ (\omega - \omega_*)^2\right], \label{eq:small_Gamma_approximation}
\end{equation}
where $\beta$ is a dimensionless quantity and $\omega_{*}$ is the frequency corresponding to $\Gamma_{*}$. Insertion of \eqref{eq:small_Gamma_approximation} and $E\approx E_*$, $dE/d\omega = (dE/d\omega)_*$ into \eqref{eq:tau_general_eqn} allows us to obtain the following analytic expression for $\tau$:
\begin{equation}
    \tau \approx 2 \left( \frac{\omega}{E} \frac{dE}{d\omega} \right)_* ~ \frac{\arctan \left( \sqrt{\beta}\right)}{\Gamma_* \sqrt{\beta}}, \label{eq:tau_approximation_quadratic}
\end{equation}
where we have chosen $\omega_{0,f} \approx \omega_*$. This expression makes manifest that the expectation $\tau \sim \Gamma^{-1}_*$ can receive important corrections due to the other terms appearing in \eqref{eq:tau_approximation_quadratic}, a feature which is also maintained in the general calculation \eqref{eq:tau_general_eqn}. Even though we will not be using this expression to estimate the lifetimes, we have nonetheless checked that the order of magnitude is still correctly predicted.

The next step is to generate data points $\{\omega_i, \Gamma(\omega_i), E(\omega_i)\}$. Recently, similar methods have been proposed \cite{Kawasaki:2019czd,Zhang:2020bec,Kawasaki:2020jnw}. In \cite{Kawasaki:2019czd,Kawasaki:2020jnw}, the authors assumed a certain profile with a single oscillation frequency and computed $\Gamma(\omega)$ by linearizing the field equation of the perturbations around such profile and computing the emitted power perturbatively in the small scalar waves. These (semi-analytic) results were then improved by the authors in \cite{Zhang:2020bec} and were benchmarked against a full numerical simulation. Given that the single frequency approximation tends to overestimate the actual $\Gamma$~\cite{Zhang:2020bec}, we choose to numerically solve the full relativistic non-linear equation of motion for $\phi$ (see Appendix \ref{appendix:numerical_time_evolution} for details and~\cite{Zhang:2020bec} for a similar approach). The decay rate $\Gamma$ is then obtained by numerically computing the outgoing flux at a position far away from the oscillon core (see Appendix \ref{appendix:computing_gamma}).

We parametrize initial conditions using
\begin{equation}
	\phi(t = 0, r) = \frac{A}{\cosh (r/\sigma)}, \quad \dot{\phi}(t = 0, r) = 0, \label{eq:initial_conditions}
\end{equation}
where $A$, $\sigma$ are free parameters that we vary. As one initializes the field according to \eqref{eq:initial_conditions} two things can happen: either the field quickly goes to the configuration $\phi = 0$ or it settles into an oscillon configuration after a mild relaxation time $t_{\text{relax}}$. Whether the former or the latter happens depends strongly on the choice of $A, \sigma$ in \eqref{eq:initial_conditions}, which is something that makes manifest both the chaotic and attractor nature of oscillons.

This is very useful because one can then trade time evolution with sampling of initial conditions. The strategy is to choose a set of $N$ different initial configurations parametrized by \eqref{eq:initial_conditions} and evolve each of them a time $t_{\text{relax}}$\footnote{The exact value for $t_{\text{relax}}$ depends on the particular initial condition and $p$.} until the oscillon configuration is found. Every initial configuration will find a different oscillon configuration, i.e. $\{A_i, \sigma_i\} \to \{ \text{oscillon}(\omega_i) \}$ after a time $t_{\text{relax}}$, with these configurations being related to each other by time evolution. All in all, this results in a huge numerical advantage since $N t_{\text{relax}} \ll \tau$. We refer the interested reader to Appendix \ref{appendix:more_details_of_relax_and_fast_forward} for a more in-depth explanation of this strategy, which we dub the \textit{relax and fast-forward} method.

\subsection{Results}
\label{sec:results}

In this section we present the numerical results that we have obtained with the numerical strategy outlined above. In particular, we give a complete picture of the lifetimes of oscillons in monodromy potentials, which has remained somewhat elusive until now, and we perform a dedicated analysis of the regime $1 - p \ll 1$.

The first result is the shape of the function $\Gamma(\omega)$ for several values of $p$, which can be found in Fig. \ref{fig:Gamma-w}. One can see that a dip appears for certain values of $p$. The tendency is to start from a shallow dip at small frequencies for negative large $p$ that becomes deeper and closer to $m$ as one approaches $p = 0$. At $p=0$ the dip is lost and $\Gamma$ tends to be a bit larger than at small negative $p$. This is followed by $\Gamma$ getting increasingly smaller as $p$ starts to get closer to $1$, recovering a dip very close to $m$. In all cases, the trajectories terminate at some frequency close to $m$ where we see the core amplitude collapse to $\phi=0$, corresponding to the death of the oscillon.

\begin{figure}[t]
\centering
\includegraphics[width=0.6\textwidth]{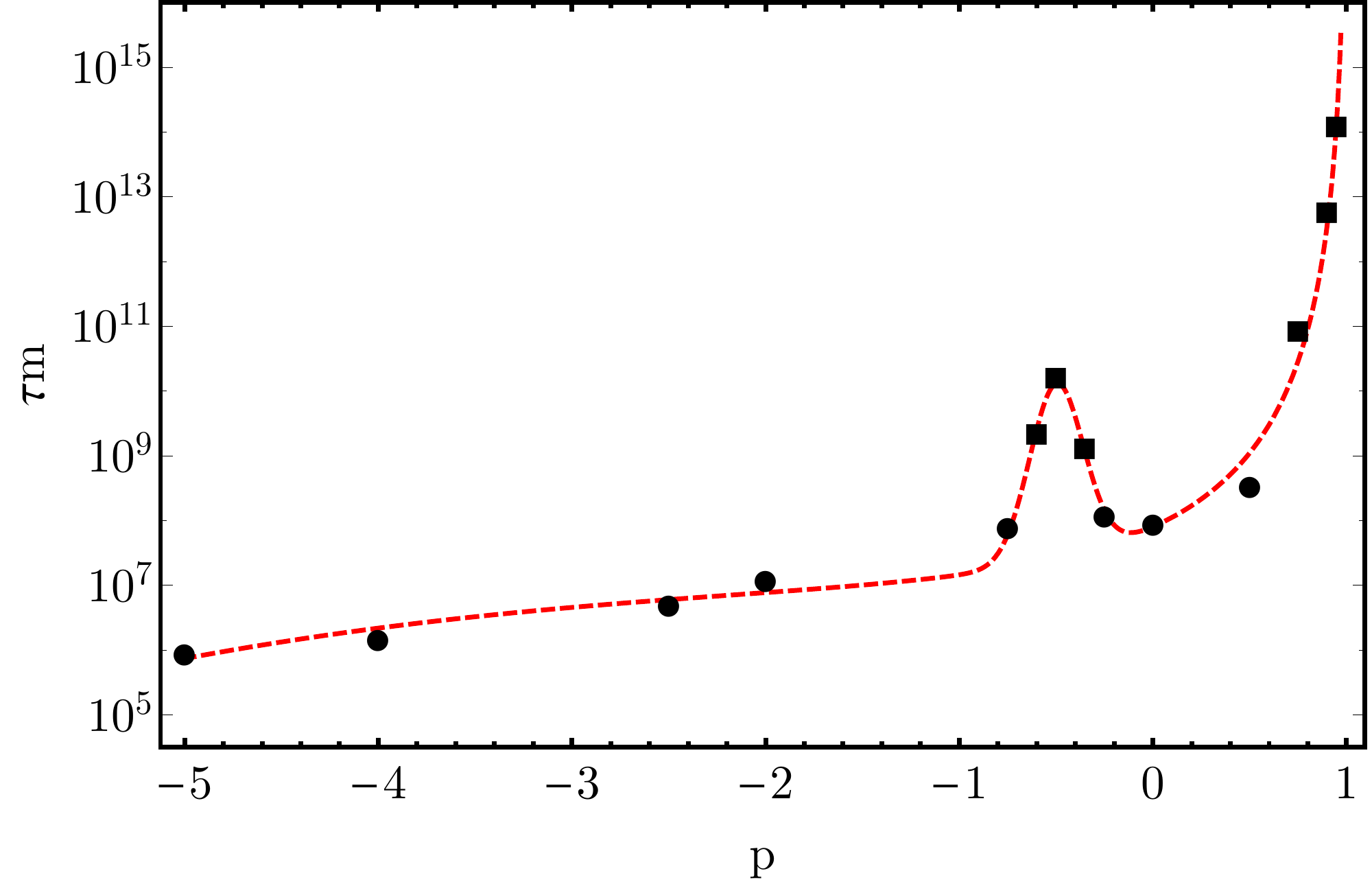}
\caption{\small Lifetime $\tau$ of oscillons that we have studied (dots) in the family of potentials~\eqref{eq:potential} as a function of $p$. Round dots have been benchmarked against an explicit numerical time evolution of the oscillon's whole lifetime, while square markers are results from the \textit{relax and fast-forward} method only. Dashed line is merely an eye guide.}
\label{fig:tau_vs_p}
\end{figure}

From the results shown in Fig.~\ref{fig:Gamma-w} we can compute oscillon lifetimes using \eqref{eq:tau_general_eqn}. Results are shown in Fig.\ref{fig:tau_vs_p}. For values of $p$ corresponding to small enough lifetimes, the \textit{relax and fast-forward} results have been benchmarked against the explicit time evolution of the corresponding oscillons during their whole lifetimes, whose values are reported in Table \ref{tabu:lifetimes_estimated_vs_real}. In particular, we find a maximum relative error of $7~\%$, which shows the quality of our method.

The lifetimes shown in Fig.~\ref{fig:tau_vs_p} follow the dependence on $p$ expected from our analysis of Sec.~\ref{sec:understanding}. In particular, the lifetime becomes insensitive to $p$  as $p \to -\infty$ and approaches $\tau \sim 10^5/m$, that is the lifetime of oscillon supported by the potential \eqref{Vtilde}. Secondly, the lifetime increases exponentially as $p \to 1$. Finally, we unexpectedly find a bump in the region $-1 \leq p \leq 0$, which appears to arise from an interplay between having a small and negative $V''(\phi)$ and being efficient in minimizing the binding potential. The power of our method is clear from our results for  $p=0.95$: we find $\tau \sim 10^{14}/m$, while following the oscillon evolution for such a huge lifetime would have taken $O(10^5)$ years using present day computers.

\begin{table}[b]
\centering
\begin{tabular}{ | c | c | c | c |}
 \hline
$p$ & $\tau m$ estimated & $\tau m$ real & Relative Error ($\%$) \\
 \hline
 $-5$  &  $8.45 \times 10^5$ & $9 \times 10^5$ & 6.11 \\
\hline
 $-4$  &  $1.39 \times 10^6$ & $1.5 \times 10^6$ & 7.33 \\
\hline
 $-2.5$  & $4.74 \times 10^6$  & $5 \times 10^6$ & 5.20 \\
\hline
$-2$  & $1.13 \times 10^7$ & $1.2 \times 10^7$ & 5.83 \\
\hline
\end{tabular}
\caption{\small Comparison between estimated lifetimes based on \eqref{eq:tau_general_eqn} and actual lifetimes for oscillons evolved with an explicit numerical time evolution (real) for some values of $p$.}
\label{tabu:lifetimes_estimated_vs_real}
\end{table}

\begin{figure}[t]
\centering
\includegraphics[width=0.47\textwidth]{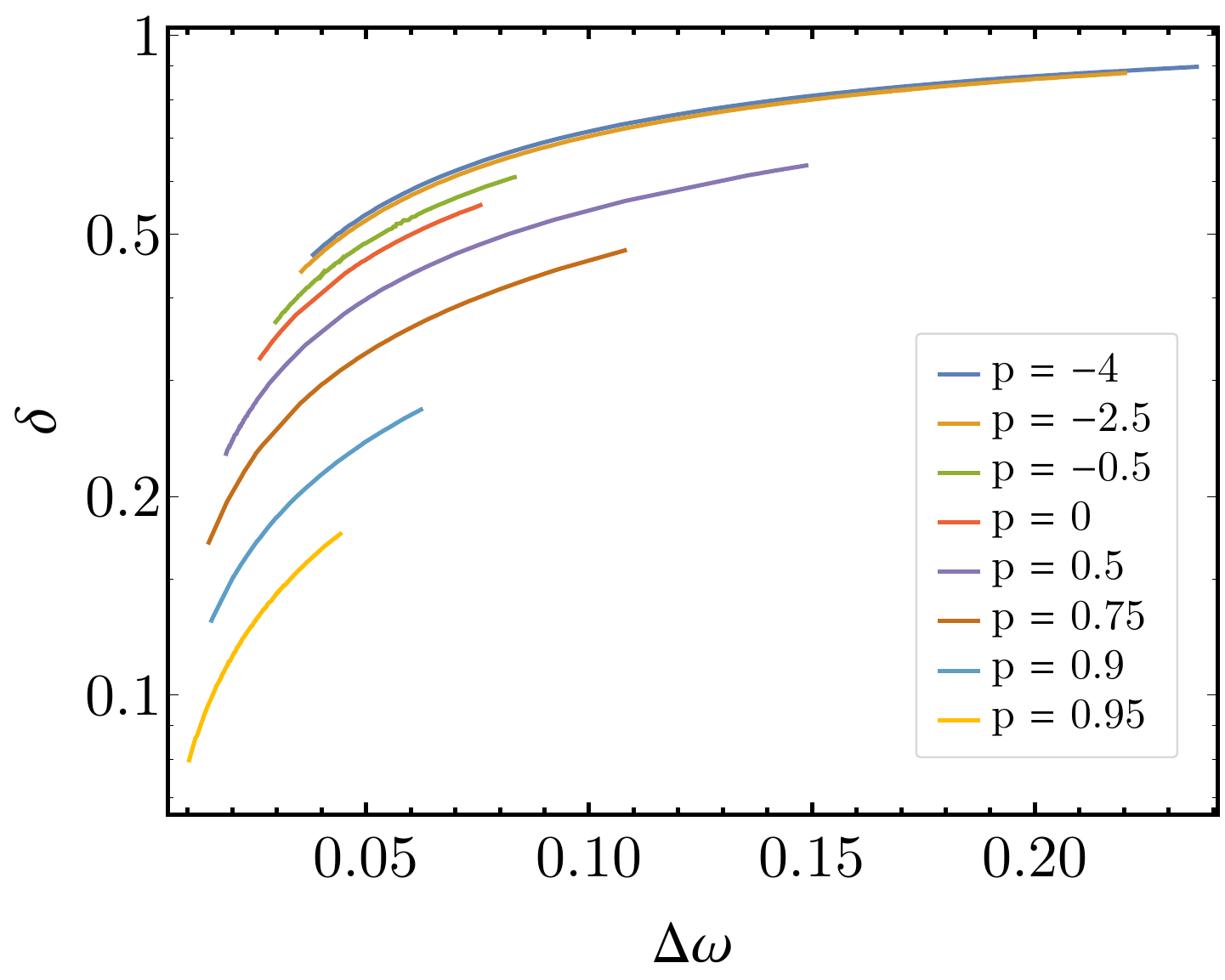} \quad
\includegraphics[width=0.47\textwidth]{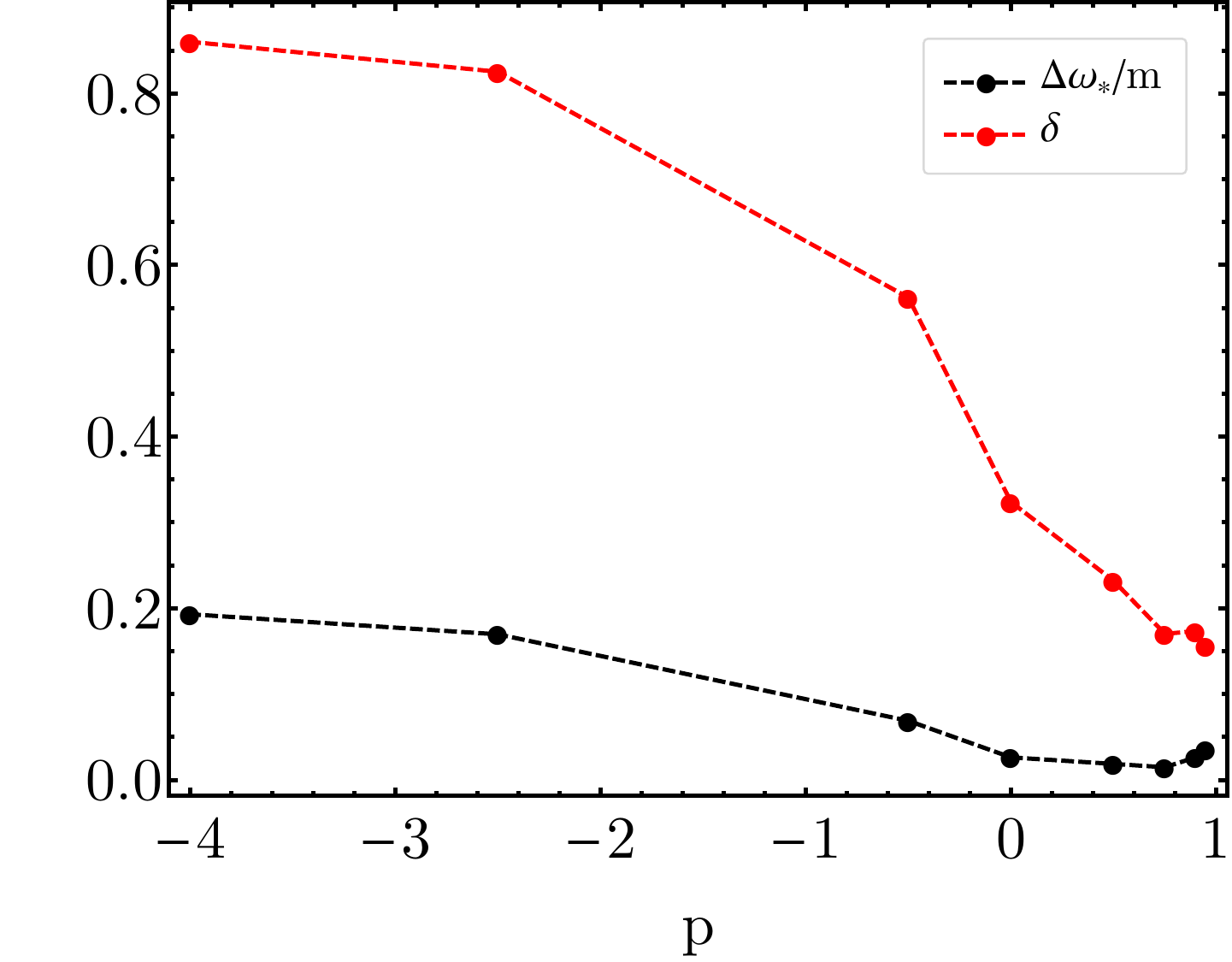} 
\caption{\small
Extraction from the numerical simulations of the binding frequency $\Delta\omega \equiv m- w_{osc}$ and the binding potential $\delta\equiv 1 - \frac{V(\phi_0)}{m^2\phi_0^2/2}$ with $\phi_0$ the amplitude of oscillation at the core. 
The left panel shows the relation between $\delta$ and  $\Delta\omega$ at each time (frequency) of the evolution. The right panel shows the  relation obtained by selecting the typical frequency $\omega_*$ where the oscillon spends most time (corresponding to the  minimum of $\Gamma(w)$ shown in Fig.\ref{fig:Gamma-w}).
}
\label{fig:delta-w}
\end{figure}

As we have argued in Section \ref{sec:understanding}, the first step to understand oscillon longevity comes from the notions of \textit{binding energy per particle}, encoded in $\Delta \omega$, and the binding potential \eqref{crit}, encoded in $\delta$. From the data of our simulations these quantities can be easily extracted and are shown in Fig.~\ref{fig:delta-w}, where we see a clear correlation between the binding frequency $\Delta \omega$ and the binding potential $\delta$. In particular, the tendency is that both $\Delta \omega$ and $\delta$ decrease as $p$ increases. Typical values in simulations for different $p$ give a range of $\delta$ from 0.1 to 0.85. This supports the criterion suggested above that $\delta$ must be a sizeable fraction of unity. Moreover, another result from Fig.~\ref{fig:delta-w} is that in the $1 - p \ll 1$ limit, $\delta$ is of order 0.1 - 0.2.

\begin{figure}[t]
\centering{}
\includegraphics[width=0.6\textwidth]{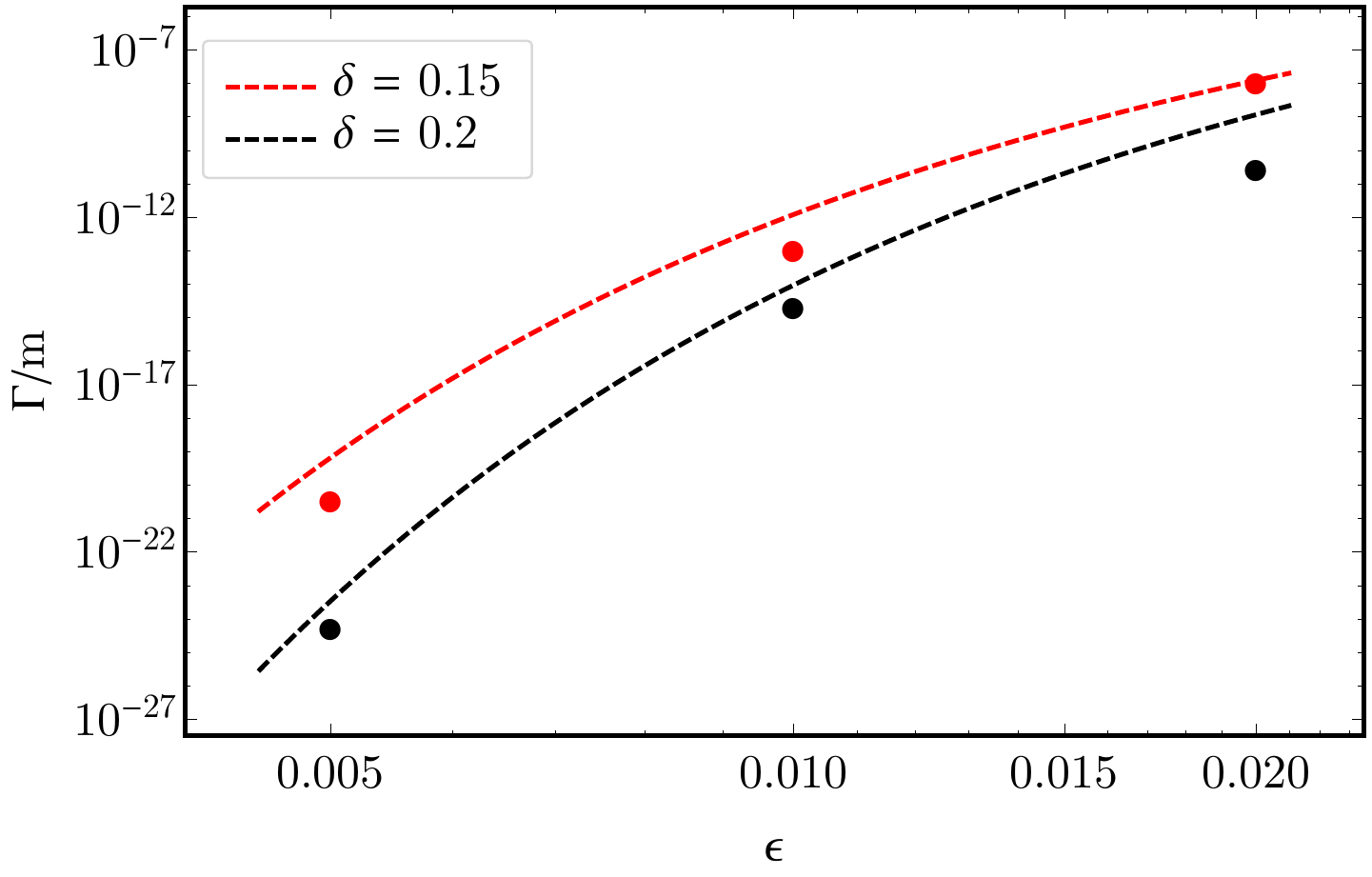}
\caption{{\small Decay rate $\Gamma$ of oscillons for $\epsilon\equiv1-p \to0$. Markers are data from numerics while dashed lines correspond to the analytic prediction \eqref{scaling}.}}
\label{fig:log_Gamma_epsilon}
\end{figure}

Finally, obtaining the $\Gamma(\omega)$ curve becomes particularly challenging for $0.95 < p < 1$, due to the fact that we must substantially increase the size of our lattice. Instead, switching to the $(\epsilon, \delta)$ language with potential \eqref{Vsmallp-1} allows us to test our analytic result \eqref{scaling} for $\Gamma$. We thus numerically solve the equation of motion with potential \eqref{Vsmallp-1} by initializing the field with a Gaussian profile with parameters determined by \eqref{gauss}. We evolve such configuration for a time $\sim (\epsilon ~ m)^{-1}$ and show the results for $\Gamma$ in Fig.~\ref{fig:log_Gamma_epsilon}. From that, we see that  $\Gamma$ is only slightly overestimated by the analytic result \eqref{scaling}, which however correctly captures the behavior with $\epsilon$. This gives further numerical evidence that the lifetime of oscillons increase exponentially as $p \to 1$.

It is instructive to look at the main numerical result  summarized in Fig.~\ref{fig:tau_vs_p}, from the perspective of the corpuscular picture. 
The lifetime  $\tau$ turns out to be significantly larger than the estimate in \ref{sec:non-exceptional} ($\tau  \sim 10^4 m^{-1}$). Monodromy potentials are expected to enhance $\tau$ from  classical field theory intuition, due to properties of the potential (flattening $V$ and not too negative $V''$). Still, the dependence of $\tau$ on $p$ seems quite remarkable. In particular there are two salient features: the extreme growth at $p\to1$ and  the peak near $p=-1/2$. The reason for the former is clear, the closeness to the $\phi^2~\log\phi$ exceptional potential. The origin of the peak is less clear to us at present, and seems more accidental. 

\begin{figure}[t]
\centering
\includegraphics[width=0.45\textwidth]{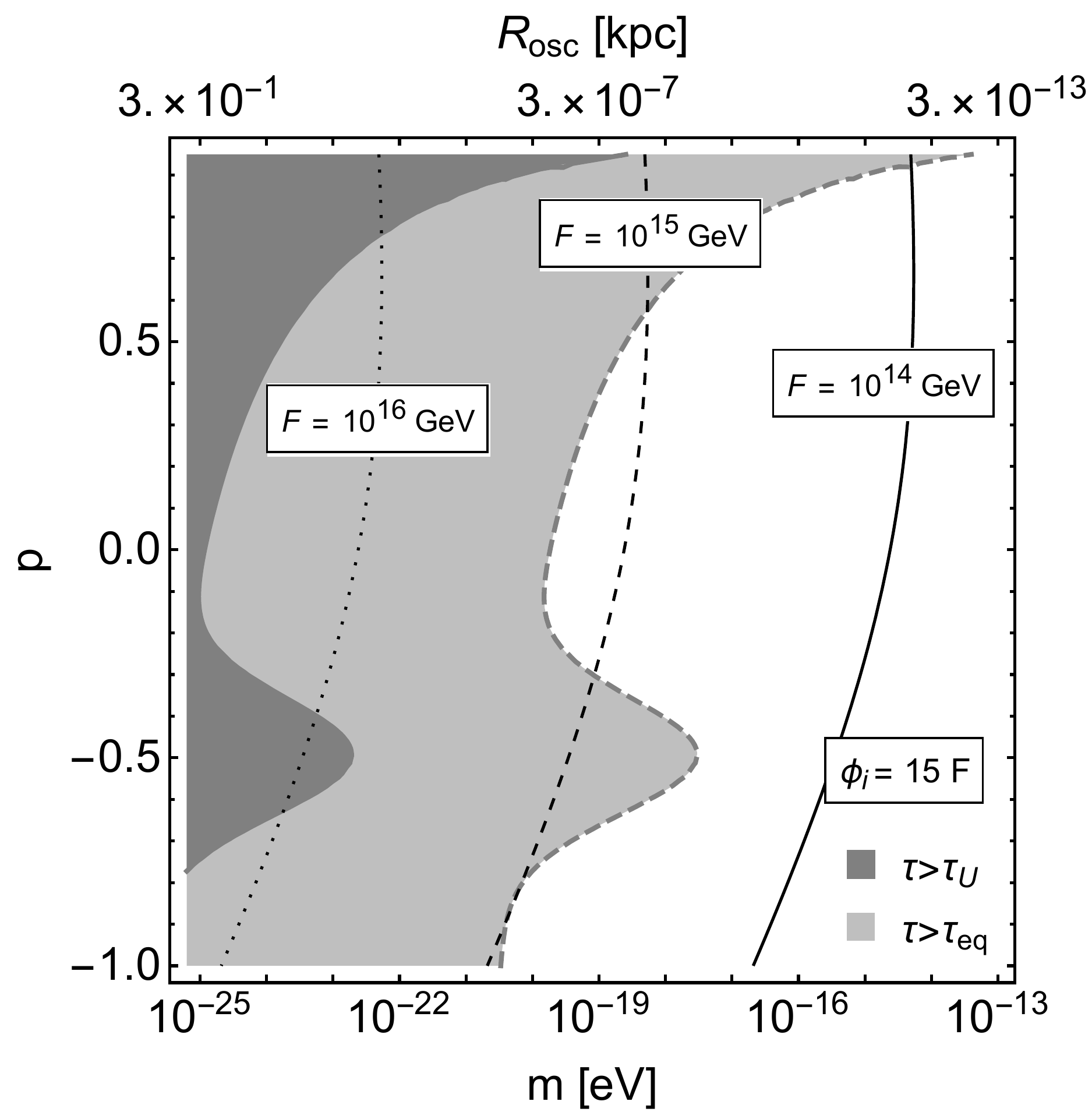}
\includegraphics[width=0.48\textwidth]{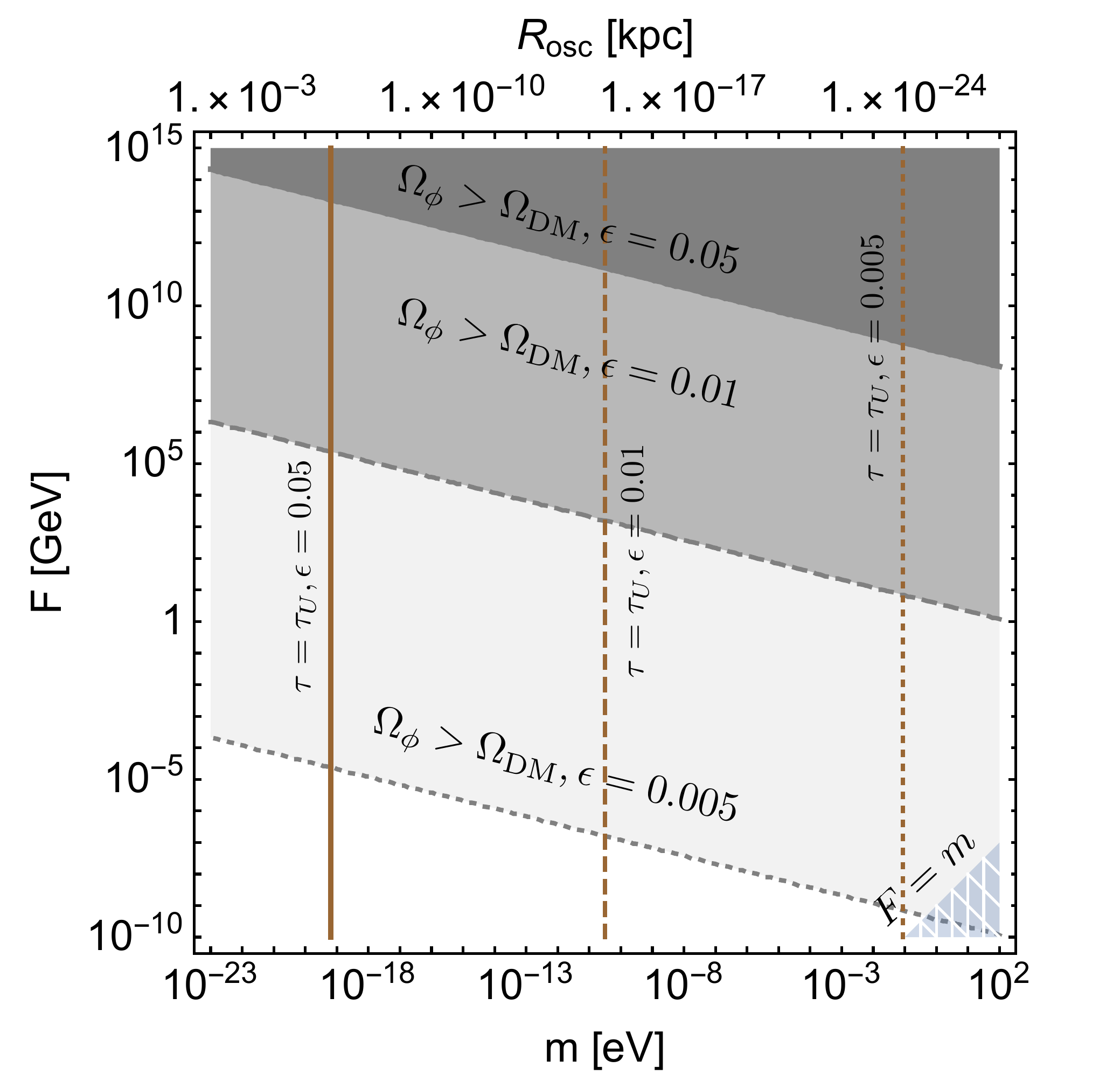}
\caption{{\small \textbf{Left:} values of $p$ and $m$ for which oscillons supported by monodromy-like potentials with $p\leq 0.95$ are stable until today (dark gray) or until matter-radiation equality (light gray). Contours of values of $F$ for which the background scalar reproduces the dark matter abundance are also shown, when the initial field value is fixed to $\phi_i=15~F$. \textbf{Right:} Constraints on $F$ and $m$ for monodromy-like potentials with $\epsilon\equiv 1-p\leq 0.05$ from dark matter overproduction. Values of the scalar mass for which oscillons are stable until today are denoted with vertical lines, which move to the right as $\epsilon$ decreases. In the lower right corner of the figure $F\sim m$ and the theory becomes strongly coupled. Close to this line, quantum annihilations in the oscillon core become important and alter oscillon lifetimes~(see e.g.~\cite{Hertzberg:2010yz, Hertzberg:2020xdn}).}}
\label{fig:pheno}
\end{figure}

\section{Discussion and Conclusions}
\label{sec:conclusions}

The observational impact of oscillons crucially depends on their lifetime. In this paper, we provided some analytical understanding for the observed extraordinary longevity of large amplitude oscillons as well as a novel numerical strategy to reliably compute their lifetimes.

We derived estimates from a simple corpuscular picture, which is valid at small coupling/large occupation numbers and admits a classical field theory limit, to qualitatively reproduce the lifetime of oscillons supported by generic potentials, such as $\sim \phi^4$ or sine-Gordon $\sim \cos{(\phi/f)}$. 
The estimates assume no cancellations amongst various possible channels (encoded in the different self-couplings), but do not exclude that these cancellations might occur.  Oscillons that are much longer lived than these estimates  are supported by ``monodromy''-like potentials, which at large field values behave as $V\sim (\text{const.}+)~\phi^{2p}$ with $p<1$. Despite several efforts in recent years, the ultimate reason for such long lifetimes and their behavior with $p$ were not fully transparent. 

In this work, we have argued that one way how these cancellations can occur is with potentials that flatten at large $\phi$ while keeping the effective mass $V''(\phi)$ as close as possible to the mass in vacuum, $m^2=V''(0)$ (with negative but small $V''(\phi)/m^2$ being tolerated). The monodromy family of potentials satisfies this property and the numerical results, summarized in Fig.~\ref{fig:tau_vs_p}, indeed show long lifetimes. This is most dramatic in the  $p\to1$ limit, as they approach the {\em exceptional} form $V_{\text{log}}\sim \phi^2\log\phi$. This logarithmic potential is known to admit eternal (non-decaying) oscillons \cite{Dvali:2002fi,Kawasaki:2015vga} so, in a way, the extreme longevity is inherited by `proximity' with this exceptional theory.

Oscillons supported by monodromy potentials with $p\rightarrow 1$ can be studied both analytically and numerically, and our analysis shows that they are extremely long-lived (see Fig.~\ref{fig:log_Gamma_epsilon}), with lifetimes many orders of magnitude larger than their cousins with $p\leq 1/2$, which were previously believed to be the most longeve oscillons \cite{Olle:2019kbo}. Moreover, we presented a new numerical method to compute oscillon lifetimes, which is particularly useful when the evolution of a single oscillon is too long to be fully simulated on a computer.

Our results can have very interesting observational implications for light scalar dark matter. Let us briefly discuss some of them, while leaving a detailed study for future work (see e.g.~\cite{Arvanitaki:2019rax} for a detailed analysis of observational consequences of oscillons in axion models). Fig.~\ref{fig:pheno} summarizes our findings for monodromy-like potentials. For $p\leq 0.95$ (left figure), we find that oscillons formed during the radiation-dominated epoch can survive until matter-radiation equality for masses as large as $m\sim 10^{-14}~\text{eV}$ when $p$ is close to unity. These overdensities may then act as seeds for the formation of structures. For $m\sim 10^{-19}~\text{eV}$ and  $p$ close to unity, we find that oscillons can actually survive until today and thus potentially constitute a part of the dark matter. We confirmed that a window of negative values of $p$ around $p=-1/2$, corresponding to plateau-like potentials, with longer lifetimes than their cousins with $0<p\leq 1/2$ exists. However, in contrast to the expectations in our previous work~\cite{Olle:2019kbo}, we found that decreasing $p$ further decreases, rather than increases, an oscillon's lifetime. The origin of this window is still unclear to us and deserves further work.

Values of the scale $F$ for which the oscillating background scalar field explains the observed dark matter abundance are also shown in the left plot in Fig.~\ref{fig:pheno} for an example choice of initial background field amplitude $\phi_i\sim 15~F$, taking into account the delay in the onset of oscillations as $p$ decreases. This choice is guided by requiring that self-resonance is efficient in the radiation dominated Universe (see e.g.~\cite{Olle:2019kbo}), so that the homogeneous field can undergo fragmentation, which would then provide initial conditions for oscillon formation.\footnote{In general, the value of $\phi_i$ required to have efficient fragmentation increases as $p$ increases. In particular, for $p\simeq 0.9$, a larger value of $\phi_i$ is required than in our example choice. Nonetheless, we find that for those large values of $p$ the required $\phi_i$ is only a $O(1)$ factor larger than in our example choice.} Oscillon masses in this case can be extremely large when $m$ is very small, the precise value depending also on $p$. In particular, for $p=-1$ we find $M>10^8 M_{\odot}$ when $m\sim 10^{-25}~\text{eV}$, whereas for $p=0.95$ we find that the same oscillon mass is obtained for $m\sim 10^{-22}~\text{eV}$. This can be used to constrain the left most region of the left plot in Fig.~\ref{fig:pheno} (see e.g.~\cite{Carr:1997cn, Carr:2018rid}), which is also further disfavored by galactic~\cite{Bar:2018acw} and Lyman-$\alpha$ observations~\cite{Kobayashi:2017jcf}. 

Nonetheless, we find that for $p\simeq -1/2$ and $p\lesssim 0.9$ oscillons with masses smaller than $10^8~M_{\odot}$ can be stable today  and contribute to the abundance of scalar dark matter when $F\sim 10^{15}-10^{16}~\text{GeV}$ and $m\sim 10^{-23}-10^{-19}~\text{eV}$. Very interestingly, this is the mass range of Ultra-Light Scalar Dark Matter, where other bound structures (held together by gravity rather than by self interactions) can form at late times. 

The observational impact of models with $p$ very close to unity, which may be motivated by other frameworks (e.g. supersymmetry), is dramatically different, since oscillon lifetimes and core amplitudes increase exponentially with $1/\epsilon=1/(1-p)$. Our results for this case are shown in the right plot of Fig.~\ref{fig:pheno}. Oscillons can then be stable until today for scalar masses up to $m\lesssim 0.1-1~\text{meV}$, when $\epsilon\sim 0.005$. In order to understand the viability of this scenario, we fix the background amplitude required to obtain these oscillons from fragmentation to be equal to the core amplitude. The right plot in Fig.~\ref{fig:pheno} then shows that very large longevity comes at the expense of a much smaller scale $F$, since otherwise the background scalar field would overclose the Universe. For $\epsilon\sim 0.01$ the mass window extends to $m\lesssim 10^{-12}~\text{eV}$ and larger values of $F\lesssim 10^4~\text{GeV}$ are allowed.

Overall, our work presents evidence that the observational implications of self-interacting (ultra)light scalar dark matter may be even richer than what has been considered so far, due to the extreme longevity of oscillons supported by certain well-motivated potentials. A comprehensive study of these consequences is a promising task for future work.

\medskip 
\section*{Acknowledgments}
We thank G. Dvali and M. Hertzberg for discussions and A. Arvanitaki, S. Sibiryakov and H. Zhang for helpful comments on the first version of this paper. The  work  of  FR  is supported in part by National Science Foundation Grant No.  PHY-2013953.
This work was partly supported by the grants FPA2017-88915-P and SEV-2016-0588 from MINECO and 2017-SGR-1069 from DURSI. IFAE is partially funded by the CERCA program of the Generalitat de Catalunya.

\appendix

\section{Numerical Time Evolution}
\label{appendix:numerical_time_evolution}

Time evolution of field configurations is generated by numerically solving the Euler-Lagrange equation of motion for $\phi$ with potential \eqref{eq:potential} and spherical symmetry,
\begin{equation}
    \partial_t^2 \phi(t,r) - \partial_r^2 \phi(t,r) - \frac{2}{r} \partial_r \phi(t,r) + V'(\phi(t,r)) = 0.
\end{equation}
It is convenient to rescale $\phi$, $t$ and $r$ so that they become dimensionless. This can be achieved by doing $\phi \to \phi/F$ and $t,r \to m t, ~ m r$. In the remainder of this section, $\phi$, $t$ and $r$ will be understood to be dimensionless.

Discretizing the equation of motion in a lattice $0 \leq r_i \leq r_{\text{max}}$ with spatial resolution $\Delta r \equiv r_{i+1} - r_i$, the field configuration at time $t + \Delta t$ is given by
\begin{align}
    \phi(t+\Delta t, r_i) = 2 \phi(t, r_i) &- \phi(t-\Delta t, r_i) + (\Delta t)^2 \left[ \frac{\phi(t, r_{i+1}) - 2 \phi(t,r_i) + \phi(t,r_{i-1})}{(\Delta r)^2} \right. \nonumber \\
    &+ \left.\frac{\phi(t,r_{i+1}) - \phi(t, r_{i-1})}{r_i \Delta r} - \phi(t, r_i)\left[1 + \phi(t, r_i)^2 \right]^{p-1} \right],
\end{align}
where we have introduced the specific form of the potential \eqref{eq:potential}. This equation is valid for all $r_i$ except the first and last points of the lattice: $r_i = 0, r_{\text{max}}$. For $r_i = 0$ we impose Neumann boundary conditions, while for $r_i = r_{\text{max}}$ we implement second order absorbing boundary conditions following \cite{Salmi:2012ta}. The typical sizes and resolutions that we use are $r_{\text{max}} = 80$, $\Delta r = 0.05 = 5 \Delta t$. We have written the code in Python.

\section{Computing $\Gamma$ numerically}
\label{appendix:computing_gamma}

Oscillons slowly decay by emitting scalar waves, so we consider the outgoing flux at a position $r =r_d$ far away from the core and larger than the oscillon size $R$,
\begin{equation}
    \Phi(t) =  4 \pi r_{d}^2 \, \dot{\phi}(t, r_d) \, \phi'(t, r_d), \label{eq:flux}
\end{equation}
where derivatives of $\phi$ are computed numerically as
\begin{equation}
	\dot{\phi}(t, r_d) = \frac{\phi(t, r_d) - \phi(t - \Delta t, r_d)}{\Delta t}, \quad \phi'(t, r_d) = \frac{\phi(t, r_d + \Delta x) - \phi(t, r_d - \Delta x)}{2 \Delta x}. \label{eq:phi_derivatives_numerically}
\end{equation}
This function will be oscillatory in time, so we will be interested in averaging it over a sufficient long period $T$, which we typically choose to be $T = N \Delta t \sim 10^2/\omega$. The decay rate $\Gamma$ will then be
\begin{equation}
    \Gamma = \frac{4 \pi r_d^2 \left|\sum_{i = 0}^{N} ~ \dot{\phi}(t_i, r_d) \, \phi'(t_i, r_d) \right|}{ N E},
\end{equation}
where $\dot{\phi}, ~ \phi'$ are computed using \eqref{eq:phi_derivatives_numerically} and $t_i = t + i \Delta t$. As for the energy $E$, it is computed using SciPy's \textit{integrate.cumtrapz} method and is also replaced by an average in the period $T$.

\section{More Details of the Relax and Fast-Forward Method}
\label{appendix:more_details_of_relax_and_fast_forward}

\begin{figure}[htpb]
\centering
  \includegraphics[width=0.55\textwidth]{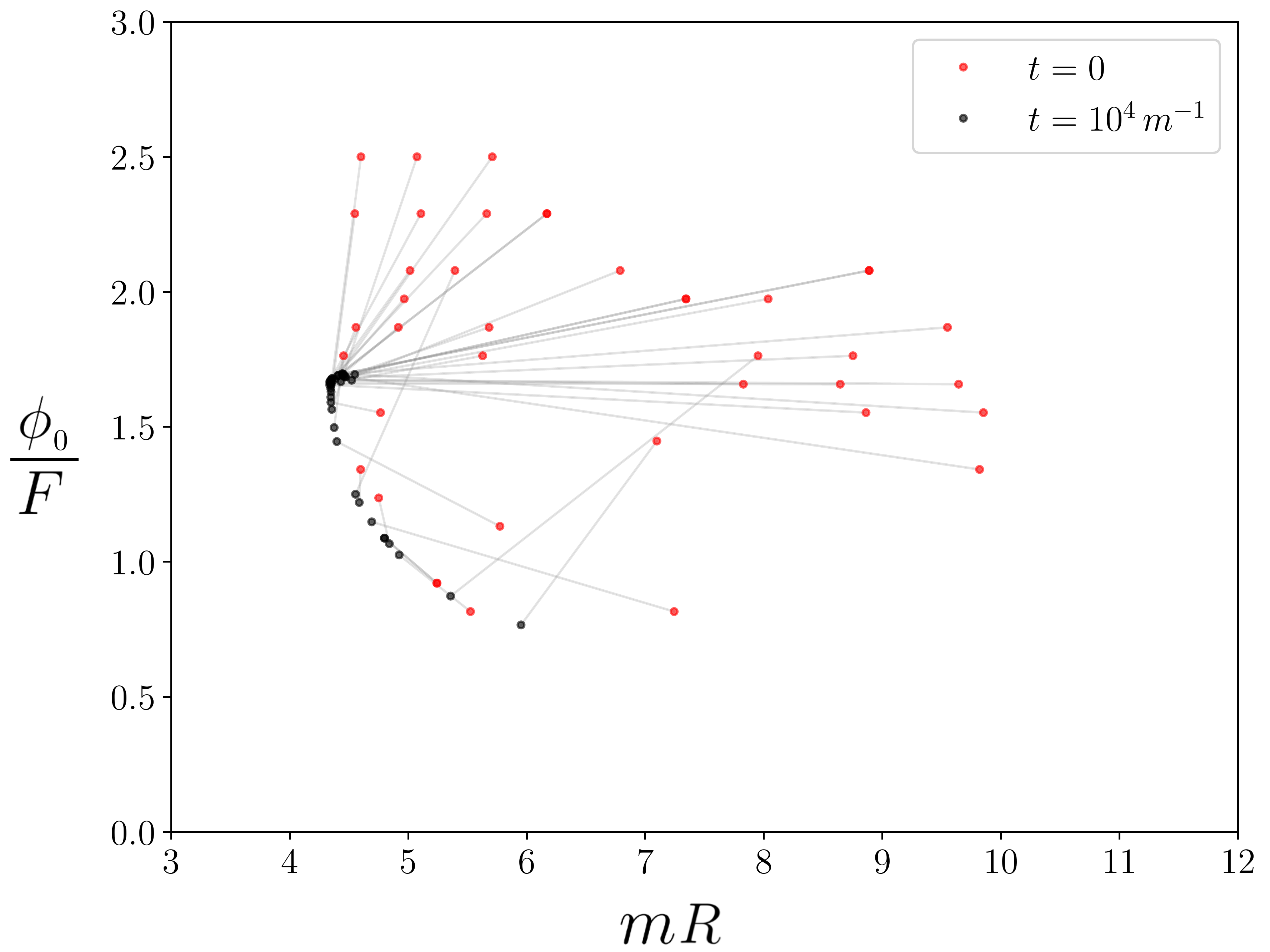}
  \caption{\small Different initial conditions evolved for a time $t = 10^4 m^{-1}$ for $p = -2.5$. One can see how different initial conditions cluster into the oscillon configuration. Lines connecting initial to final states are meant to be for identifying purposes, they are not the actual trajectories in parameter space. Here $R$ is numerically computed as the position where $90~\%$ of the energy in the lattice lies.}
  \label{fig:attractor}
\end{figure}

The working principle of this method is that one can buy computing time from time evolution by sampling initial conditions. This is only an advantage if the relaxing time to reach the oscillon configuration when starting from a somewhat random initial configuration is much smaller than the total oscillon lifetime, $t_\text{relax} \ll \tau$. This happens due to the attractor nature of oscillons. As one can see in Fig. \ref{fig:attractor}, several different initial conditions as parametrized by  \ref{eq:initial_conditions} cluster in parameter space after a mild relaxation time. 

\begin{figure}[htpb]
\centering
  \includegraphics[width=0.75\textwidth]{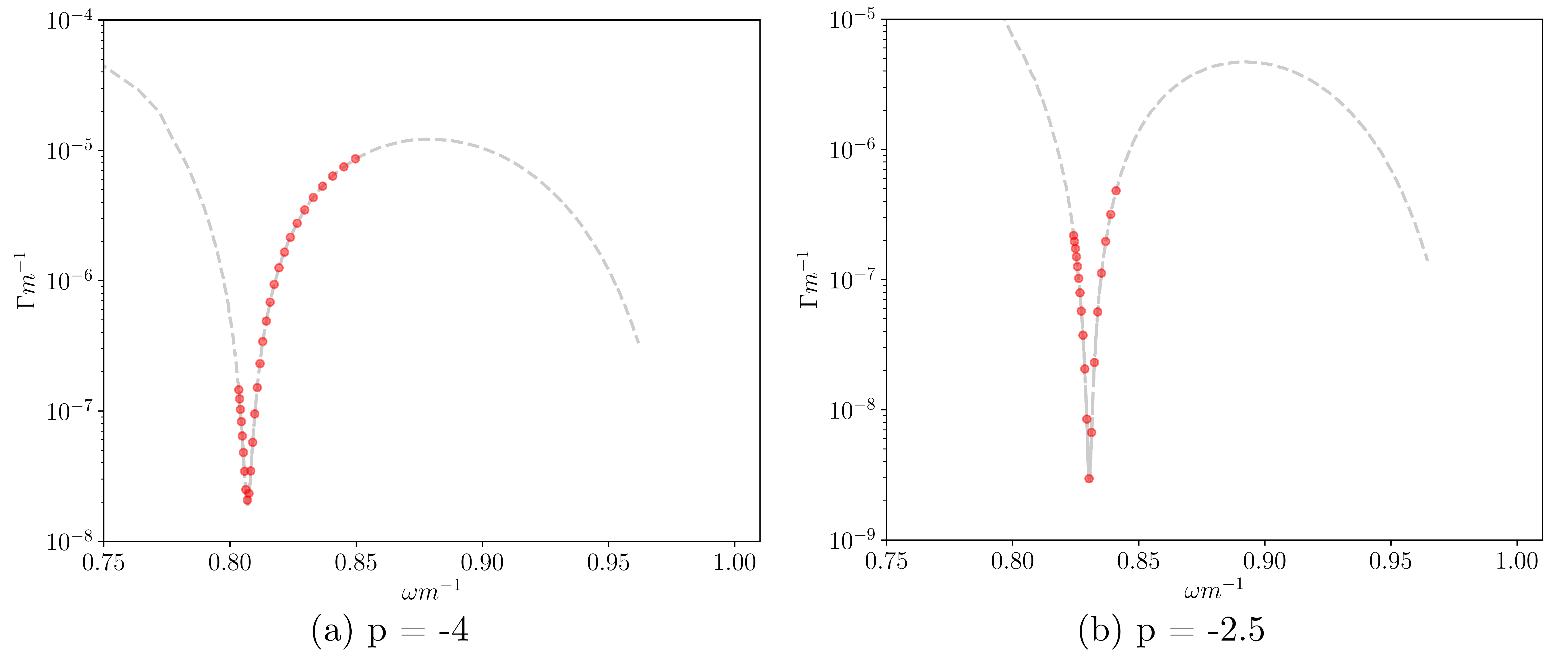}
  \caption{\small Comparison between points generated by the \textit{Relax and Fast-Forward} method (red dots) and the full time evolution (black dashed). Each red dot comes from a \textit{different} initial condition that is evolved for a time $t_\text{relax} = 10^4 m^{-1}$, while the black dashed lines are generated from a single initial condition by applying numerical time evolution.}
  \label{fig:tm_vs_full}
\end{figure}

The second important observation is that the different configurations that are found after time $t_\text{relax}$ are all part of the same oscillon and are related by time evolution. This is illustrated in Fig. \ref{fig:tm_vs_full}, where we see that different initial conditions all find the oscillon trajectory after a mild relaxation time $t_\text{relax} = 10^4 m^{-1}$ at different configurations (different values of $\omega$). 

\begin{figure}[htpb]
\centering
  \includegraphics[width=0.55\textwidth]{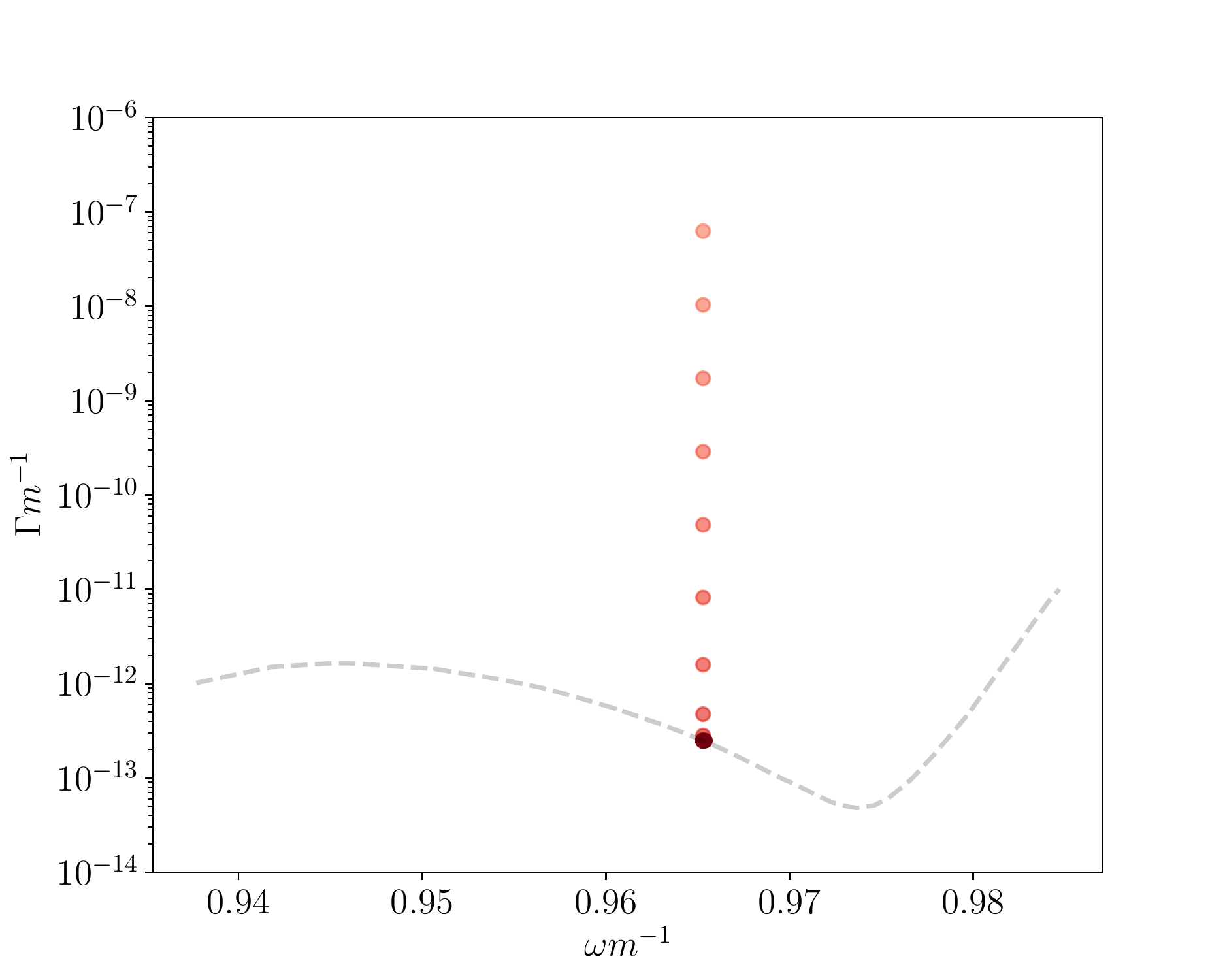}
  \caption{\small Example of relaxation of a particular initial condition for $p=0.9$. Increasingly darker markers determines the direction of time. After a period of rapid evolution, the configuration ends in the attractor oscillon configuration.}
  \label{fig:single_ic_example_p0d9}
\end{figure}

In some cases, the lifetimes will be so large that the oscillon trajectory in parameter space will not be easily computable. What this means is that there will be some $p$'s for which the black dashed curves of Fig. \ref{fig:tm_vs_full} will not be known, and by exploring several initial conditions we will have to reconstruct them. This problem reduces to checking that the relaxation time is large enough, and this can depend not only on $p$, but also on the initial condition. Nevertheless, once one sees that a configuration barely moves in parameter space after a period of relatively fast movement, we can be sure that the configuration has relaxed into the oscillon. This is illustrated for a particular initial condition of $p = 0.9$ in Fig. \ref{fig:single_ic_example_p0d9}, where we have added the reconstructed trajectory.

\bibliography{biblio}
\bibliographystyle{JHEP.bst}

\end{document}